%% file: trend50.tex
\documentclass[preprint,12pt]{elsarticle}
\usepackage{amssymb}
\usepackage{amsmath}
\usepackage{caption}
\usepackage[nolist]{acronym}
\usepackage[displaymath, mathlines]{lineno}
\usepackage{color}
\journal{Astroparticle Physics}
\acrodef{TREND}{TianShan Radio Experiment for Neutrino Detection}

\DeclareMathOperator\erf{erf}

\begin{document}
\begin{frontmatter}
\title{Autonomous radio detection of air showers with the TREND50 antenna array}
\author[subatech]{D.~Charrier}
\author[vub]{K. D.~de Vries}
\author[ihep]{Q.~Gou}
\author[naoc]{J.~Gu}
\author[ihep]{H.~Hu}
\author[naoc]{Y.~Huang}
\author[naoc]{S.~Le Coz\corref{cor1}}
\ead{lecozsandra@yahoo.com}
\author[naoc,lpnhe]{O.~Martineau-Huynh\corref{cor1}}
\ead{omartino@in2p3.fr}
\author[lpc]{V.~Niess\corref{cor1}}
\ead{niess@in2p3.fr}
\author[naoc]{T.~Saugrin}
\author[plata]{M.~Tueros}
\author[naoc]{X.~Wu}
\author[naoc]{J.~Zhang}
\author[ihep]{Y.~Zhang}

\cortext[cor1]{Corresponding author}
\address[subatech]{SUBATECH, Institut Mines-T\'el\'ecom Atlantique, CNRS-IN2P3, Universit\'e de Nantes, Nantes, France.}
\address[vub]{Vrije Universiteit Brussel, Physics Department, Brussels, Belgium}
\address[ihep]{Institute of High Energy Physics (IHEP), Chinese Academy of Sciences, Beijing, P.R. China}
\address[naoc]{National Astronomical Observatories of China (NAOC), Chinese Academy of Science, Beijing, P.R. China.}
\address[lpnhe]{Sorbonne Universit\'e, Universit\'e Paris Diderot, Sorbonne Paris Cit\'e, CNRS,
Laboratoire de Physique Nucl\'eaire et de Hautes Energies (LPNHE), Paris, France \\
}
\address[lpc]{Universit\'e~Clermont Auvergne, CNRS/IN2P3, LPC, Clermont-Ferrand, France}
\address[plata]{Instituto de F\'isica La Plata-CONICET, La Plata, Argentina}

\begin{abstract}
TREND50 is a radio detection setup of 50 self-triggered antennas working in the 50-100\,MHz frequency range and deployed in a radio-quiet valley of the Tianshan mountains (China). TREND50 achieved its goal: the autonomous radiodetection and identification of air showers. Thanks to a dedicated offline selection algorithm, 564 air shower candidates were indeed selected out of $7\cdot10^8$ transient radio signals recorded during the 314 live days of data taken during the first two years of operation of this setup (2011 and 2012). This event rate, as well as the distribution of the candidate directions of arrival, is consistent with what is expected from cosmic-ray-induced air showers according to simulations, assuming an additional $\sim$20\% contamination of the final sample by background events. This result is obtained at the cost of a reduced air shower detection efficiency, estimated to be $\sim$3\%.
This low efficiency is mostly due to the large amount of dead time of the setup. This result
paves the way for the GRANDProto35 experiment, the first stage of the GRAND project.
\end{abstract}
\end{frontmatter} 

\input{intro}
\input{setup}
\input{calib}
\input{data}
\input{selection}
\input{results}

\input{conc}

\section{Acknowledgments}
We thank the staff of the Ulastai Radio Observatory for their continuous effort in setting up and maintaining the TREND setup through the years. We thank Frank G. Schr\"oder, Charles Timmermans and Mauricio Bustamante for their usefull comments on the text. This work was supported by the France China Particle Physics Laboratory, the Key Projects of Frontier Science of Chinese Academy of Sciences (grant QYZDY-SSW-SLH022), the Strategic Priority Research Program of Chinese Academy of Sciences (grant No.XDB23000000), the National Key R\&D Program of China (grant No.2018YFA0404601), the Natural Science Foundation of China (grants No. 11105156,  11135010, 11375209, 11405180 and 11505213), the Chinese Ministry of Science and Technology and the Youth Innovation Promotion Association of Chinese Academy of Sciences. K.D. de Vries'work is supported by the Flemish foundation of scientific research (grant FWO-12L3715N). The simulations presented in this paper were produced at IN2P3/CNRS computing center, the GRIF grid site and the IHEP computing center, in particular through the France-Asia Virtual Organisation.

\section{Bibliography}
\bibliographystyle{elsarticle-num}
\bibliography{biblio}

\appendix
\input{annex}

\end{document}

%% file: intro.tex
\section{Introduction}
Radiodetection is a maturing technique for the detection of air showers (see~\cite{Huege:2016veh,Schroder:2016hrv} for reviews). The low cost of the detection units makes it an attractive solution for large detectors such as GRAND\,\cite{FangGRANDICRC:2017}, but taking full advantage of the radio technique requires that excellent performances can be achieved with a standalone radio array on air-shower detection efficiency, background rejection and primary particle reconstruction. 

The Tianshan Radio Experiment for Neutrino Detection (TREND) already obtained encouraging qualitative results on this matter with 6- and 15-antennas setups~\cite{Ardouin:2011}. Others\,\cite{CODALEMA:2017} have also proved that good reconstruction of air showers can be achieved from radio data only. Several radio setups also detected autonomously cosmic rays in the very favorable radio environment of Antarctica~\cite{Hoover:2010qt,Barwick:2016}, with an excellent rejection of background events. 

Here we report on the results of the 50-antenna stage of the TREND experiment, focusing in particular on the issue of air shower identification over an ultra dominant background. A preliminar analysis on this topic was presented in \cite{LeCoz:2017ICRC}. After introducing the setup, calibration and dataset (sections~\ref{section:setup} to~\ref{section:data}), we will detail the offline algorithm that was developed to discriminate air showers from background, and validate its performance on a simulated dataset (section~\ref{section:selection}). We will finally apply this algorithm to the experimental dataset in section~\ref{section:results}, where we will also discuss the air-shower detection performances of the TREND setup and analysis chain.

%% file: setup.tex
\section{The TREND50 experiment}
\label{section:setup}

\subsection{The TREND50 layout}
\begin{figure}[t!]
\begin{center}
\includegraphics[width=\textwidth]{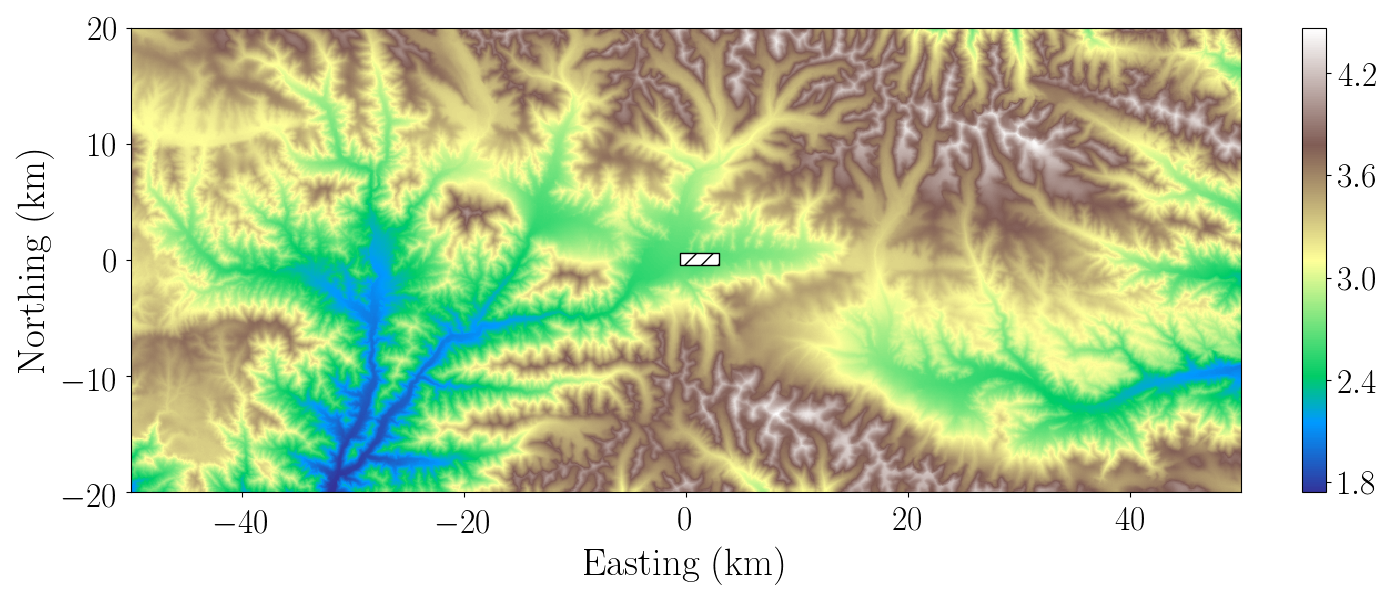} \\
\includegraphics[width=\textwidth]{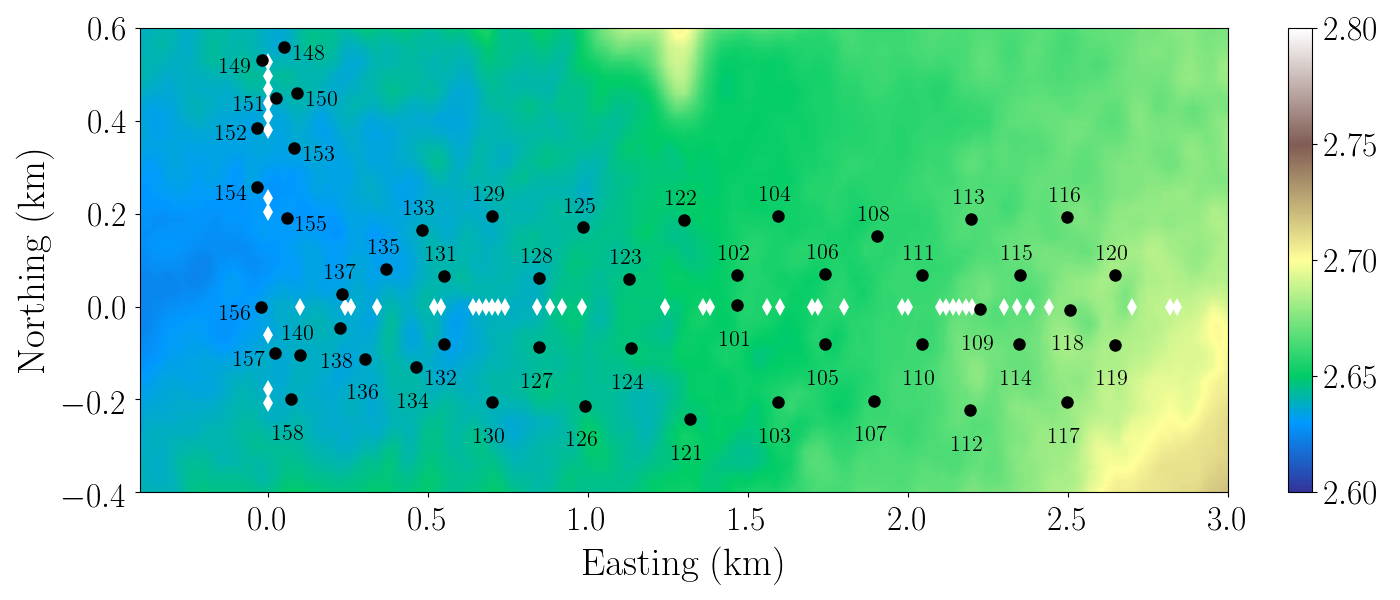}
\end{center}
\caption{the TREND50 layout. The top figure is a topographic view of the setup surrounding. It was obtained from SRTMGL1 data\,\cite{SRTMGL1} projected to UTM coordinates with the TURTLE library\,\cite{TURTLE:GitHub}. The bottom figure is a zoom on the area surrounding the layout, indicated by a hatched box on the top figure. Black dots correspond to positions of the TREND50 detection units and white diamonds to positions of the 21\,CMA $pods$ (see text for definition).
The color bar on the right represents the altitude above sea level, in km. Position are given here in the TREND referential, with x-axis along the West-East direction, y-axis along South-North, and origin taken at the cross-point of the two 21\,CMA baselines.}
\label{fig:layout}
\end{figure}

The TREND50 detector is composed of 50 radio detection units separated by a distance of 175\,m on average, thus covering a total ground surface of 1.5\,km$^2$ (see Fig.~\ref{fig:layout}). It is an upgrade of the 15-antenna setup presented in~\cite{Ardouin:2011}, and referred as TREND15 in the following. The TREND setup is deployed at the crosspoint of two valleys of the Tianshan mountains, in the Xinjiang Autonomous Region, China. This remote mountain site benefits from excellent electromagnetic background conditions, with a measured power density very close to the minimal level expected in the 50-100\,MHz frequency range \cite{Ardouin:2011}.

The TREND experiment is located at the site of the 21\,CMA radio-interferometer~\cite{21CMA}, and relies in particular on its optical fiber network to transfer data to the DAQ room (see section~\ref{section:DAQ} for details). This implies that TREND50 detection units had to be deployed at a reduced distance from the two baselines of the 21\,CMA interferometer, hence its peculiar layout.


\subsection{The TREND50 detection unit}
\label{section:unit}
\begin{figure}[t!]
\begin{center}
{\includegraphics[width=\textwidth]{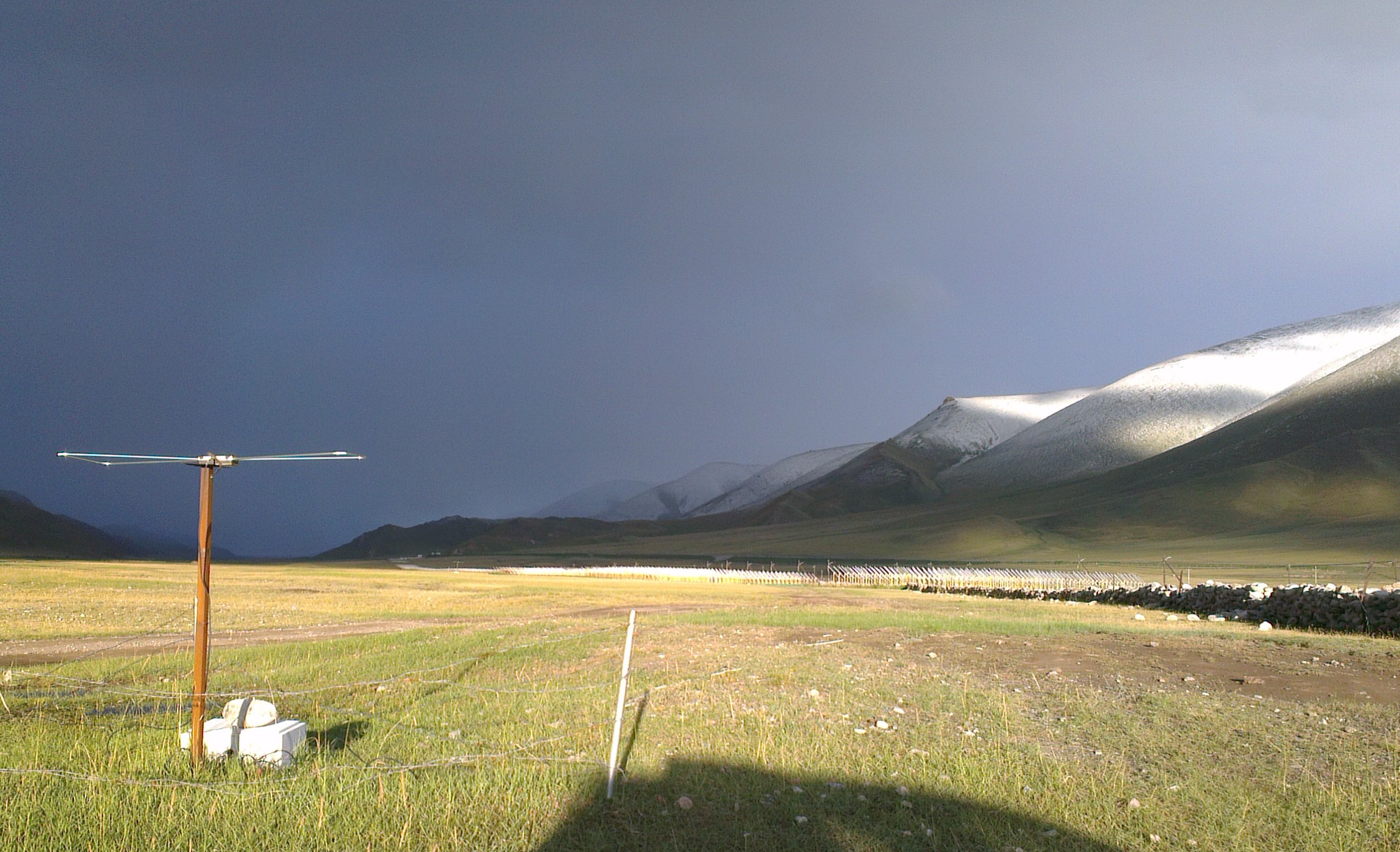}}
\end{center}
\caption{a TREND50 detection unit, composed of a bow-tie antenna and a RF transformer installed at the antenna feed-point. The front-end electronics (see text) is installed in the box placed at the foot of the detection unit. Several 21\,CMA pods -where the TREND50 signals are fed to the optical fiber network- can also be seen in the distance, at the foot of the snowy Tianshan mountains. }
\label{fig:antenna}
\end{figure}
TREND50 detection units (see Fig.~\ref{fig:antenna}) are composed of two elements :
\begin{itemize}
{\item a single-polarization bow-tie antenna, 130\,cm long and 50\,cm wide, installed horizontally at a height of 150\,cm above ground.}
{\item a RF transformer of primary/secondary impedance ratio equal to~1.5 installed at the feed-point of the antenna, with primary on the antenna side.}
\end{itemize}

\begin{figure}[t!]
\begin{center}
{\includegraphics[width=\textwidth]{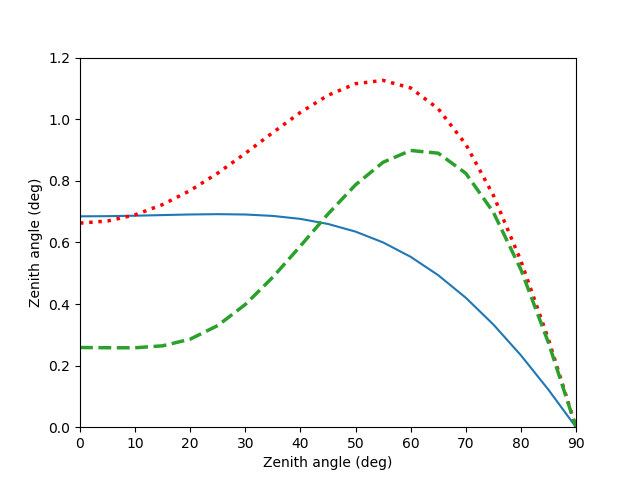}}
\end{center}
\caption{magnitude of the effective length for the TREND50 detection unit as a function of zenith angle, for an incoming monochromatic plane wave with polarization in a plane parallel to the antenna main axis, of frequency 50 (solid blue), 80 (dashed green) and 100\,MHz (dotted red). In this NEC2 simulation, the detection unit is connected to a 75\,$\Omega$ load, and an infinite flat ground is assumed, with relative dielectic constant $\epsilon_r = 10$ and conductivity $\sigma = 10^{-3}$\,$\Omega ^{-1}\mathrm{m}^{-1}$ (labeled as "sandy dry ground" conditions in NEC2).}
\label{fig:leq}
\end{figure}

In order to characterize the response of the detection unit to electromagnetic waves, the {\it antenna equivalent length} $l_{eq}$ is used. We recall that it is a complex vectorial quantity allowing to compute the voltage $V_L$ measured at the input of the electric circuit loading the detection unit in response to an electric field $\vec{E}$ propagating in the $(\theta,\phi)$ direction, through:  \\
\begin{equation}
\label{eq:leq}
V_L = \vec{l}_{eq}(\theta,\phi) \cdot \vec{E}(\theta,\phi).
\end{equation}
The parameters $\theta$ and $\phi$ are the zenith and azimuth angles of the wave direction of origin. In the TREND conventions, the zenith angle is measured following the standard cosmic-ray convention (0$^\circ$ for origin at the zenith, 90$^\circ$ for origin along the horizontal) while azimuth is measured counter-clockwise with respect to the North direction. The TREND50 detection unit was designed so that $\vec{l}_{eq}$ is maximal in the 50-100\,MHz frequency range. This was checked through simulation with the NEC2 numeric code~\cite{NEC2}, as illustrated in Fig.~\ref{fig:leq}.

\subsection{The TREND50 DAQ system}
\label{section:DAQ}
The TREND15 DAQ has been previously described in~\cite{Ardouin:2011}. The TREND50 DAQ follows the same structure, except for the trigger logic which has been upgraded. Let us first recall the main steps and then we will expand on the trigger logic. To start with, the output voltage of the TREND50 detection unit is shaped by a front-end analog chain located at the antenna foot. This electronic chain has been assembled from commercial components matched to an impedance of 75\,$\Omega$. It performs a bandpass filtering of the signal in the 50-100\,MHz range, as well as low-noise pre-amplification. The preamplification gain is 64\,dB for detection units closer than 250\,m from a 21\,CMA {\em pod}\footnote{A pod is the detection unit of the 21\,CMA radio-interferometer. It is composed of 127 LPDA antennas~\cite{21CMA}.}, 84\,dB for larger distances. Then, a coaxial cable routes the resulting voltage to the closest 21\,CMA pod where it is transduced and sent to the DAQ room ---distant by 2\,km at most--- through an optical fiber. \\

In the DAQ room, the signals of each antenna are passed to a dedicated processing unit (PU). The signals of all antennas are digitized in parallel by dedicated 8-bits ADCs integrated in the PU and running synchronously at a rate of 200 MSamples/s. For each antenna, the digitized samples are buffered in time frames of 1.342\,s and analysed on the fly by the PU, looking for local triggers (T0). The T0 algorithm is the same as for the TREND15 setup. It proceeds as follows:
\begin{enumerate}[(i)]
        \item{For each {\it time frame} we compute the mean, $\mu$, and standard deviation, $\sigma$, of the samples.}
        \item{The time frame is then divided in 262\,144 {\it subframes} of 1024 samples, each corresponding to a waveform duration of $\sim5.1\,\mu$s.}
        \item{For each subframe the maximum deviation, $d_\textrm{max} = \max(|s_i - \mu|)_{i \in [0,1023]}$, and its corresponding time, $t_\textrm{max}$, are computed.}
        \item{If $d_\textrm{max} \geq N \sigma$ a T0 trigger is issued with trigger time $t_\textrm{max}$.}
\end{enumerate}
$N$ is a scalar adjusted by the user between 6 or 8, depending on the electromagnetic background conditions at the start of the run.

The T0 triggers generated within one subframe are collected by a master unit MU. If at least four T0 triggers are causally coincident, then the MU sends a T1 signal back to the corresponding PUs, triggering the storage of the $5.1\,\mu$s waveforms participating in the coincidence to a local disk, after they are centred on $t_\textrm{max}$. The set of waveforms associated to a T1 trigger compose a so-called {\it event}.

Formaly, the T1 coincidence condition is defined as following:
\begin{equation}
|t_i-t_j| \leq \frac{d_{ij}}{c} \times T
\end{equation}
where $t_i$ and $t_j$ are the T0 time values for two antennas $i$ and $j$ (after correction for their respective signal propagation delay, see section~\ref{section:timecalib}), $d_{ij}$ the ground distance between the two antennas, $c$ the velocity of light and $T$ a factor introduced to allow for possible discrepancies between T0 and the "true" trigger time (that is, the actual time at which the electromagnetic wave reaches antenna $i$). A value $T$ = 1.1 is chosen in the data acquisition, which is a safe factor considering the TREND timing resolution (see section~\ref{section:timecalib}). \\

The TREND50 DAQ system was built with spare elements of the 21\,CMA setup, with the minimization of the DAQ dead-time as a main concern. Since the TREND15 setup described in~\cite{Ardouin:2011}, the T1 trigger was moved from offline to online in order to reduce the data volume (by a factor $\sim$10) and consequently simplify the offline analysis.
However, it should be noted that if  one time frame is processed in more than 1.342\,s (i.e. if the T1 search is slower than the data flow), then the present time frame is lost, together with the following one. In order to avoid such situation, the number of T0s recorded by a PU was limited to a maximum of 256 per subframe ($\sim$190\,Hz). Subsequent T0s from this PU are lost. As a result, the overall dead time of one PU is typically 10-15\,\% for standard noise conditions to 40-50\,\% in noisy cases, though some detection units close to a noise source might be completely blind at some occasions. In any case, the T0 and T1 rates, as well as the T1 processing times, are logged into the system for each and every time frame, allowing for a reliable offline monitoring of the DAQ livetime.

%% file: calib.tex
\section{Calibration \& source reconstruction}

\subsection{Time calibration}
\label{section:timecalib}

\begin{figure}[t!]
\begin{center}
{\includegraphics[width=6.5cm]{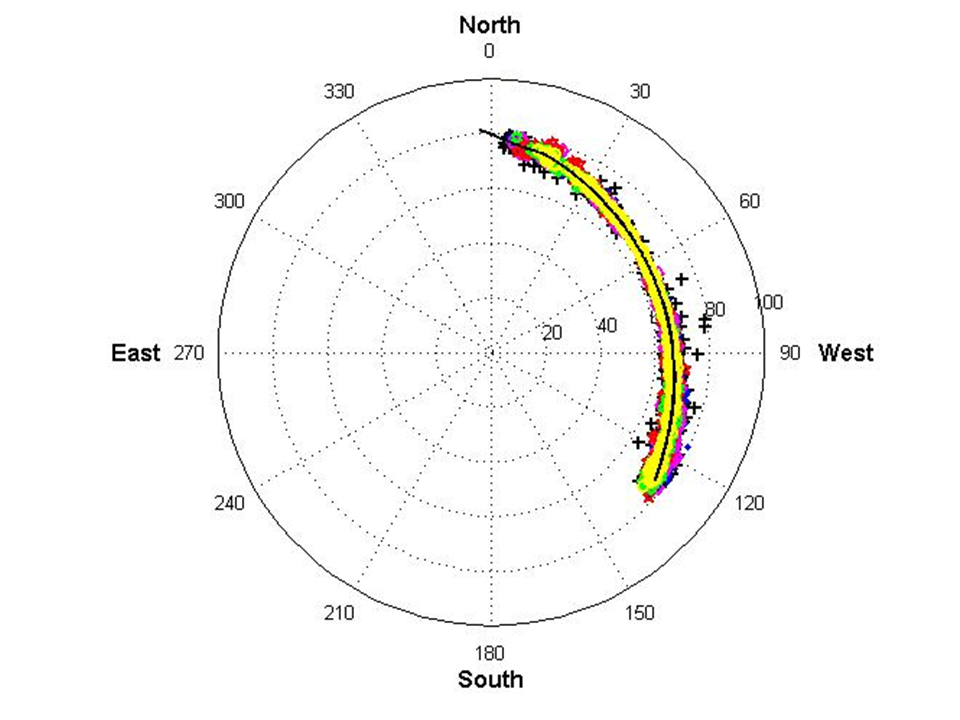}
\includegraphics[width=6.5cm]{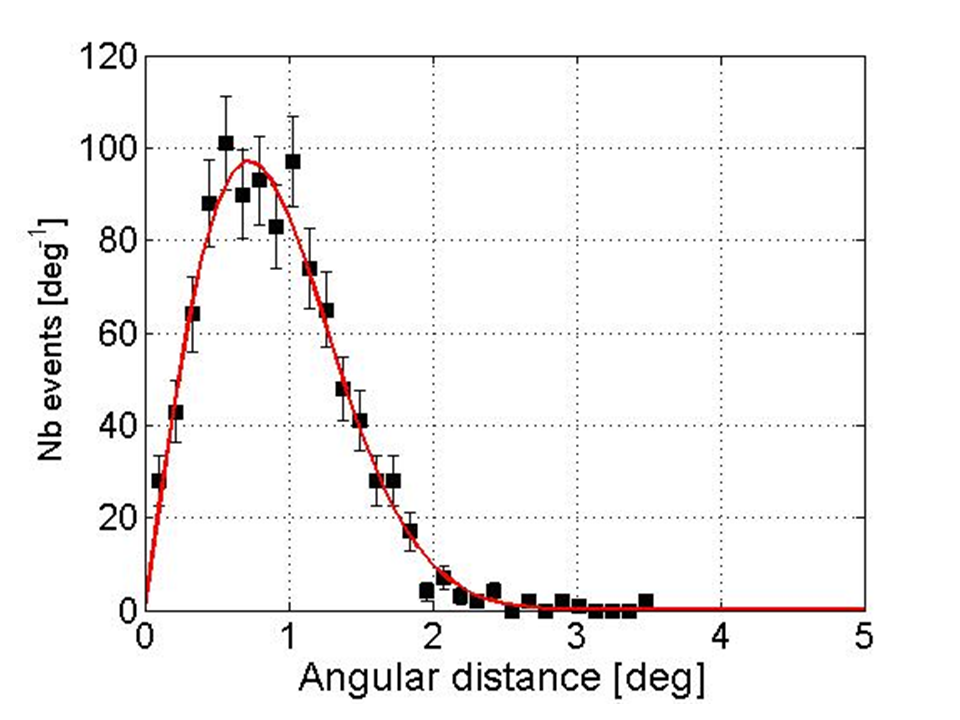}
}
\end{center}
\caption{left: skyplot of 3037 events detected within 4 minutes by the TREND50 array and reconstructed under a plane wave hypothesis. These events were most likely triggered by a plane flying above the detector. The black line is the mean reconstructed direction of the events, assumed to correspond to the trajectory of the plane.  Yellow color stands for events with multiplicity $L\geq 22$. Right: angular distance $\psi = \sqrt{(\phi-\phi_{track})^2+(\theta-\theta_{track})^2}$ between reconstructed source directions ($\theta$,$\phi$) and the plane track ($\theta_{track}$,$\phi_{track}$) for events with multiplicity $L\geq 22$. The $\psi$ distribution is fitted by the function $f(\psi) = A\exp(\frac{-\psi^2}{2\sigma^2})\sin\psi$, yielding an angular resolution $\sigma = 0.7^{\circ}$ for this airplane. }
\label{fig:track}
\end{figure}

As detailed in the previous section, signals from the detection units are digitized at the processing units through ADCs running synchronously. Time calibration in TREND50 thus simply consists of measuring the time delays induced by signal propagation inside the coaxial cables and optical fibers connecting the detection units to the DAQ room.

This calibration was performed right after array deployment by simultaneously feeding a 1~Hz square signal in the cables of two close-by detection units and measuring the time delay between the associated pulses recorded on disk. Repeating this measurement for all neighbouring antenna pairs sucessively allowed to determine the relative cable delay for all detection units. The systematic precision achieved through this method was estimated to be $\sim$10\,ns~\cite{Ardouin:2011}.

Timing calibration directly impacts the reconstruction of the direction of origin of TREND events, which is performed following a method presented in section~\ref{section:recons}. As already shown in \cite{MartineauHuynh:2012vj}, tracks of planes flying above the array could be used to evaluate the achieved angular resolution (see  Fig.~\ref{fig:track}). It highly depends on the multiplicity of the event and the zenith angle, but values are typically 2-3$^{\circ}$ for 4 or 5-fold events, and around 1$^{\circ}$ or better when multiplicity is above 10 detection units. This is enough for the TREND purpose of cosmic ray identification.

\subsection{Amplitude calibration}
\label{section:ampcalib}
%
%

Absolute calibration of the response of a radio detector is a complex measurement, in particular because calibrating sources can hardly achieve a precision much better than 10\% on the amplitude of the emitted electromagnetic field~\cite{Aab:2017lby}, and also because ground reflections induce additional uncertainties on the received signal for open-field calibration procedures. Furthermore, it was observed that the response of the TREND50 setup varies over time (see Fig.~\ref{fig:gain}), requesting a continuous monitoring of its gain.

Several radio experiments managed to perform succesfull absolute calibration of radio antennas at the $\sim$20\% precison level, or slightly better \cite{Aab:2017lby,Tunka:2015, Nelles:2015gca, LOPES:2015eya, 2012JInst...7P0011A}, also demonstrating good agreement between experimental results and modelisation of the antenna response performed with the NEC code. We therefore rely in our analysis on the NEC simulation of the detection unit response, restricting our calibration procedure to the DAQ chain.  \\

Fortunately, a continuous monitoring of its status can be performed thanks to a stable, periodic and well-understood calibration source, which, atop of that, comes for free: the diffuse radiation from our Galaxy. It has been extensively studied and modeled (see \cite{Gal:2016} and reference therein for instance), and when folded with the antenna response, it yields a reliable estimate of the signal at the output of the detection unit~\cite{2008ICRC....5..921L}. Ratio of the signal measured at the output of the DAQ chain to this expected value at input hence directly provides a measurement of the full DAQ chain gain. This method, used to calibrate the TREND DAQ chain, is detailed in the following.

\subsubsection{Voltage level at the output of the TREND50 detection units}
An unpolarized radiation of spectral radiance $B_{\nu}$ induces a power spectral density $\mathcal{P}$ at the output of a detection unit given, in units of V$^2$/Hz, by:
\begin{equation}
\label{eq:v2sky}
\mathcal{P}(\nu) = \frac{R_L}{2} \int_{4\pi} B_{\nu}(\theta,\phi,\nu)A_{eff}(\theta,\phi,\nu)\sin \theta d\theta d\phi \end{equation}
where $R_L = 75$\,$\Omega$ the input impedance of the electric circuit loading the TREND50 detection unit. The factor $\frac{1}{2}$ arises from the projection of the unpolarized radiation on the antenna arm and $A_{eff}(\theta,\phi,\nu)$ is the effective area of the TREND50 detection unit. It is related to its equivalent length, defined in Eq.~\ref{eq:leq}, through:
\begin{equation}
A_{eff} = \lvert \vec{l}_{eq} \rvert^2 \frac{Z_0}{R_L}
\end{equation}
where $Z_0$ is the impedance of free space~\cite{Balanis}. Note here that impedance mismatch effects as well as RF transformer insertion loss ($f_{ins}$=0.43\,dB) are included in the computation of $A_{eff}$. \\

\begin{figure}[t!]
\begin{center}
{\includegraphics[width=\textwidth]{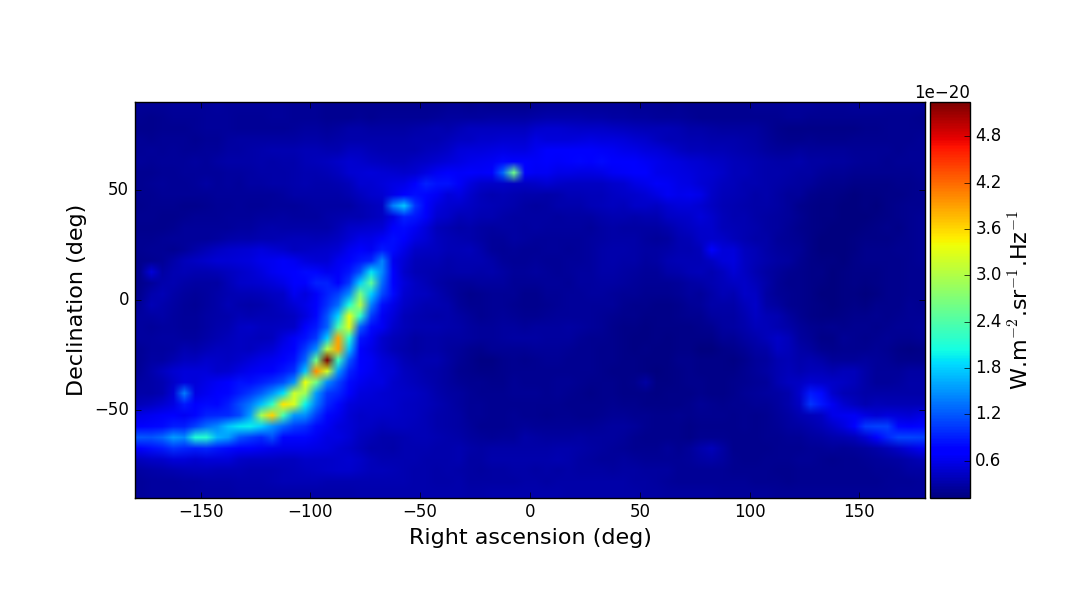}
}
\end{center}
\caption{spectral radiance of the sky $B_{sky}$ at 75\,MHz in equatorial coordinates. This map was generated with the LFmap code (www.astro.umd.edu/$\sim$emilp/LFmap), an alternative to GSM.}
\label{fig:skyb}
\end{figure}

The main source of radiation to be considered in the 50-100\,MHz frequency range is the sky diffuse emission. Models such as GSM~\cite{Gal:2016} allow generating sky maps of $B_{sky}(\theta,\phi)$ (see Fig.~\ref{fig:skyb}) through a wide range of frequencies with an accuracy better than 1\%, thus allowing to compute $\mathcal{P}_{sky}$, the corresponding power spectral density at DAQ input through Eq.~\ref{eq:v2sky}.

Note here that the transit of the Galactic plane through the antenna field of view induces a characteristic variation of $\mathcal{P}_{sky}$ with time. At the latitude of the TREND experiment, a minimum is expected around 11\,h in local sideral time (LST) before the power increases to roughly twice this value around 18\,h LST (see Fig.~\ref{fig:gain}). This periodic time variation incidentally provides an efficient way to identify malfunctioning antennas. \\

The radiation by ground also induces a stationary noise at the antenna output. In the 50-100\,MHz frequency range, this can be modeled as a blackbody radiation of temperature $T_{ground}$ = 290\,K. Using the Rayleigh-Jeans approximation:
\begin{equation}
B_{ground}=\frac{2 \nu^2 k_B T_{ground}}{c^2}
\end{equation}
the associated power spectral density can be computed from Eq. \ref{eq:v2sky}, using also the antenna effective area computed in free space and limiting the integral to the 2$\pi$ steradian facing ground. The resulting value was found to be negligeable --- 7dB below the sky contribution at 100\,MHz --- and this will therefore not be detailed here.

Neglecting other sources of radiation, we finally consider the noise level expected at the output of the TREND detection unit simply equal to the sky contribution: $\mathcal{P}_{ant}=\mathcal{P}_{sky}$.


\subsubsection{Gain computation}
\label{gaincomp}
During data acquisition, the Power Spectrum Density (PSD) at DAQ output is continuously estimated for all 50 detection units over calibration time frames of duration $\Delta t_\mathrm{cal}$, typically set to 20\,min. To this end, time traces of $N=512$ samples are randomly selected with a stationary probability, $p$. This probability, $p$, is set such that on average 10,000 traces are selected in every calibration time frame. This random sampling procedure was used because it provides a non periodic sampling over the whole calibration time frame while not disturbing the main DAQ task. Selecting all traces during $\Delta t_\mathrm{cal}$ would have blocked the DAQ, since it is too CPU consuming for our system. Subsequently, the Fourier transforms of the traces are computed with a Fast Fourier Transform (FFT) algorithm and a Hann windowing using the IPPS library. Averaging the squared amplitudes yields the PSD estimate for the time frame, in units of V$^2$/Hz:
\begin{equation}
PSD(\nu) = \frac{2 N K^2}{F_s f_{Hann}} \frac{1}{M}\sum_{i=1}^M{|FFT_i|^2(\nu)}
\label{eq:psd}
\end{equation}
where the factor 2 accounts for the fact that the IPPS FFT is computed for positive values of frequencies only. The coefficient N at the numerator arrises from the forward normalization convention we used for IPPS FFT. In addition, $K=0.0076$ is the TREND's ADC conversion coefficient in V/ADC count, $M$$\sim$10,000 is the actual number of selected traces, $F_s$=200\,MHz is the ADC sampling frequency and $f_{Hann}=0.3743$ is a normalisation factor accounting for the Hann windowing. 
\\

For every calibration time frame, the average gain of each detection unit DAQ chain, $\overline{G}$, is estimated in three steps:  \\

\begin{enumerate}[(i)]

\item{First, we estimate the frequency dependent voltage gain, $G_{i}$, for a set of discrete frequency values $\nu_i \in [55, 95]$\,MHz in steps of 5\,MHz, as:
\begin{equation}
G_{\nu_i}=\sqrt{\frac{PSD(\nu_i)}{\mathcal{P}_{ant}(\nu_i)}}
\label{eq:gv}
\end{equation}
}

\item{ Then, in order to filter out sharp band emitters, we remove outliers whose gain values depart more than 3\,$\sigma$ from the mean value of the set, where $\sigma$ is the estimated standard deviation from the set. }  \\

\item{  Finally, the average gain $\overline{G}$ is estimated as the mean of the subset values, as:
\begin{equation}
\overline{G} = \frac{1}{N_I} \sum_{i \in I}{G_{\nu_i}}
\end{equation}
where $I$ is the subset of valid values and $N_I$ its cardinality.  \\
}
\end{enumerate}

\begin{figure}[t!]
\begin{center}
{\includegraphics[width=6.5cm]{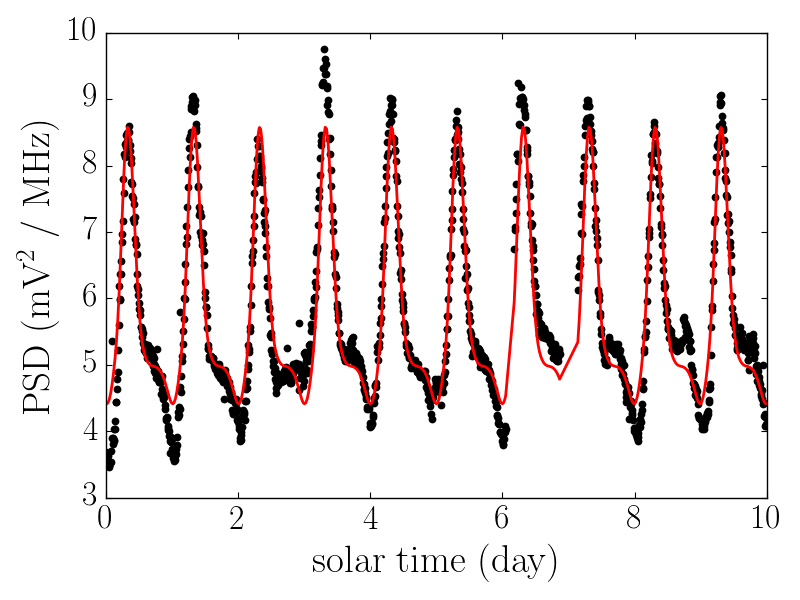}
\includegraphics[width=6.5cm]{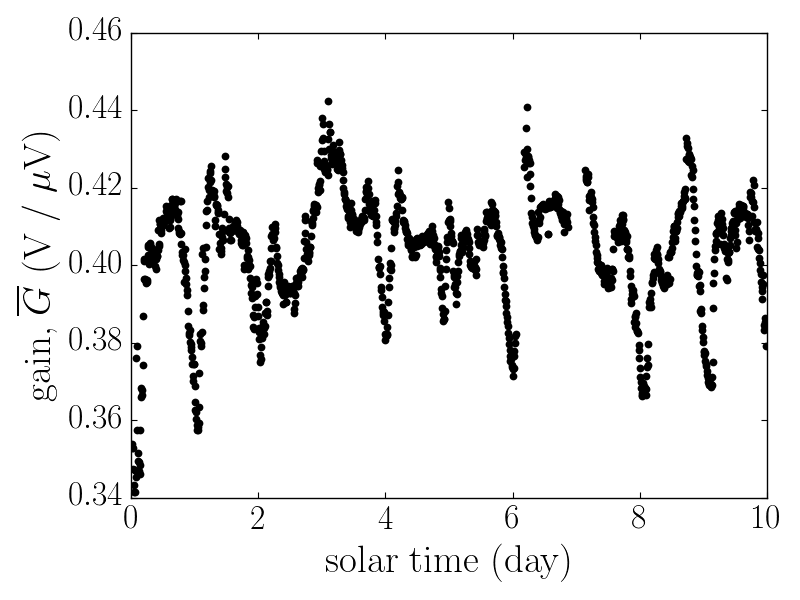}
}
\end{center}
\caption{left: the black dots stand for the measured terminal power density, $PSD(\nu_i=75\,\mathrm{MHz}$), as a function of local sideral time for antenna 113 during runs R3036 to R3049 taken from August 25$^\mathrm{th}$ to September 4$^\mathrm{th}$, 2011. Superimposed (red line) is the expected power density $\mathcal{P}_{ant}(\nu_i=75\,\mathrm{MHz})$ at output of the detection unit. Right: the corresponding estimated gain, $\overline{G} $.}
\label{fig:gain}
\end{figure}

An illustration of the whole procedure is given in Fig.~\ref{fig:gain}. Note however that our calibration procedure does not allow to distinguish a noisy period affecting all frequencies in the 50-100\,MHz range from a variation of the gain. Nevertheless, such noisy periods can be partially vetoed during the analysis, as explained in section~\ref{section:cuts}.

\subsubsection{Systematic effects}
\label{load}

\begin{figure}[t!]
\begin{center}
{\includegraphics[width=\textwidth]{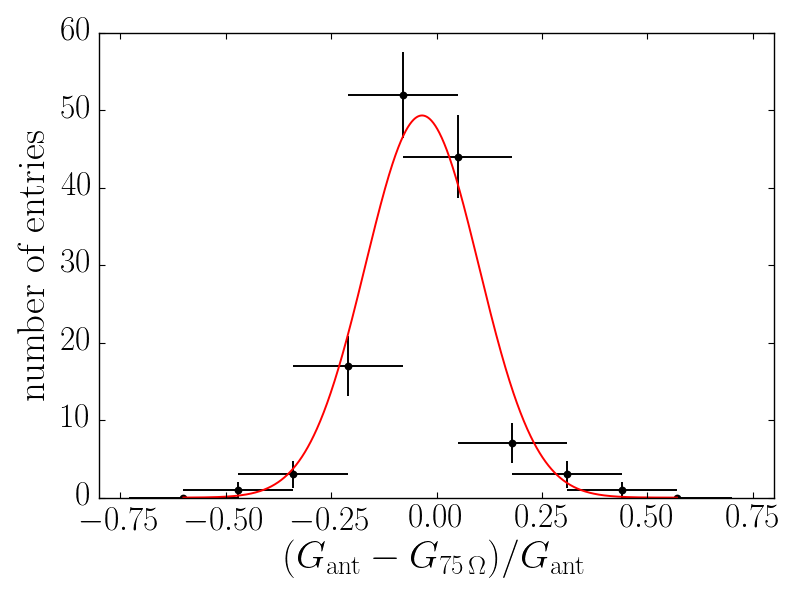}  
}
\end{center}
\caption{distribution of the relative difference $R_G = \frac{G_{ant}-G_{75\Omega}}{G_{ant}}$ between gains computed with antenna and 75\,$\Omega$ plug at DAQ input for the 128 valid measurements performed. We find $<R_G> = -3.6 \pm 1.2$\,\%. This result validates the standard DAQ chain calibration method and in particular the hypothesis that background sources beyond the sky emission have a minor contribution to the statonnary noise level at the output of TREND detection units. A 13\% precision on the DAQ calibration can also be infered from the standard deviation of the $R_G$ distribution. \\
}
\label{fig:load}
\end{figure}

Several effects may affect the calibration method above detailed: the electromagnetic background could be larger than the minimal level given by the sky, the actual antenna response could differ from the modeled $A_{eff}$ value, or the NEC "sandy dry" model (see Fig.~\ref{fig:leq} for details) considered in our treatment may not represent the true ground conditions. These effects may bias the actual value of $\mathcal{P}_{ant}$ and consequently the gain. \\

In order to quantify these effects and the validity of the TREND50 gain computation, the detection units were occasionally replaced by resistive end plugs ($R_L$ = 75\,$\Omega$). Since the end plug impedance matches the DAQ input impedance, the noise level at the DAQ input can be reliably estimated by its Johnson-Nyquist noise:
\begin{equation}
\label{eq:v75}
\mathcal{P}_{75\Omega}=k_B T R_L
\end{equation}
in units of V$^2$/Hz, where $T = T_{amb} + T_{LNA}$, with $T_{amb}$ the physical temperature of the end plug, taken equal to 290\,K, and $T_{LNA}$ = 28\,K the nominal value of the noise equivalent temperature of the low noise amplifier. Amplifier noise indeed has a non-negligeable contribution here, whereas sky temperature of several hundreds Kelvin justify that it was neglected in the computation of $\mathcal{P}_{ant}$.
\\
Computing the PSD value for this setup with Eq.~\ref{eq:v2sky}, and taking the square root of its ratio to the expected noise level at input computed with Eq.~\ref{eq:v75}, directly yields a measurement of the DAQ chain voltage gain. \\

Given the burden of this calibration procedure --- plugging end plugs into electronic chain inputs instead of detection units and back ---, the gain values were computed with this method for a total of 128 measurements only, recorded on different antennas and at different moments of the TREND50 operation period. They were then compared to gain values determined through the standard procedure, performed right after the detection units were reconnected. The relative difference between the two gain values is displayed in Fig.~\ref{fig:load}. The relative difference between the gains computed through these two independent methods has a mean value of 3.6\%, thus validating the calibration procedure of the TREND50 detection units above detailed.

%% file: data.tex
\section{The TREND50 data}
\label{section:data}
\subsection{Data set}
\begin{figure}[t!]
\begin{center}
{\includegraphics[width=\textwidth]{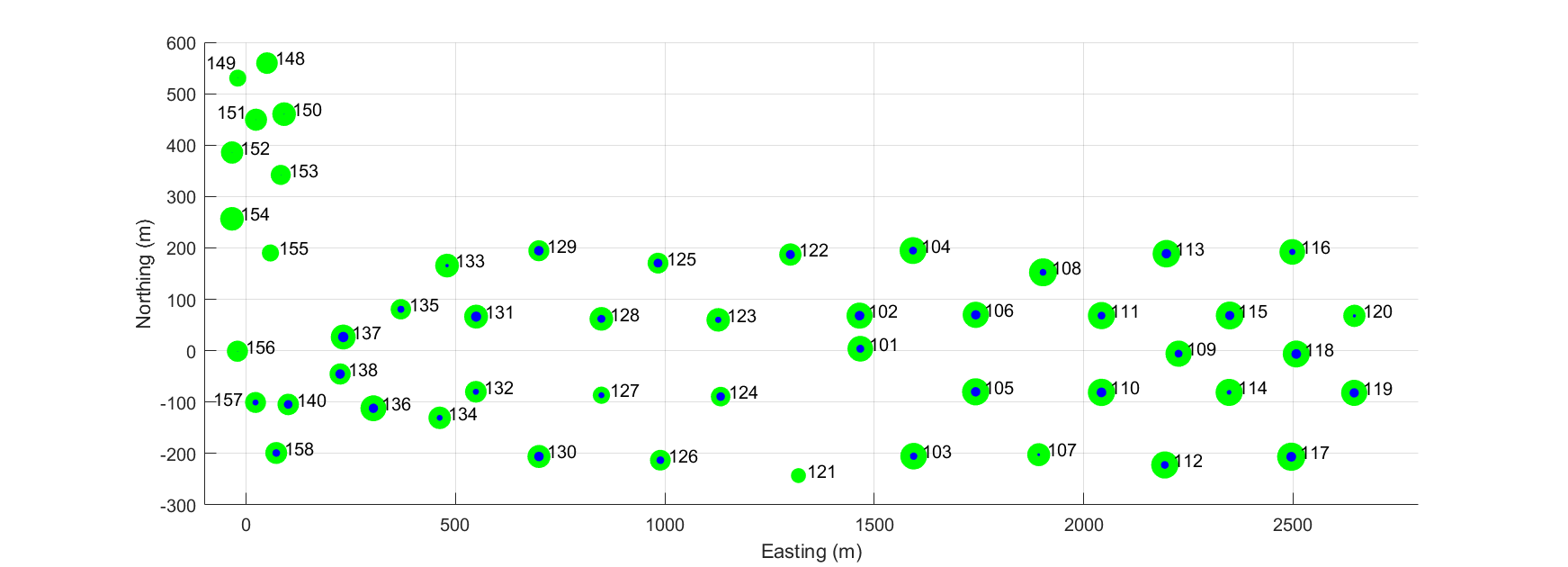}  
}
\end{center}
\caption{DAQ live times of the TREND antennas represented by circles with a radius proportionnal to their values. Top: East-West period, bottom: North-South period. The outer circle represents the total livetime of each period (314.3 and 120.6 days respectively). The average up time fraction of the detection units is 66 and 46\% for each of the two periods.}
\label{fig:livetime}
\end{figure}

Two data sets are used in the analysis presented here: the first one was recorded between January 13, 2011 and December 6, 2012, while antenna arms were oriented along the East-West axis. This data set corresponds to a total of 314.3 live days, during which $8.6\cdot10^{10}$ T0 and $7.3\cdot10^{8}$ T1 triggers were processed and $9.7\cdot10^9$ timetraces recorded.
Antennas were then rotated towards the North-South axis --- except for antennas 148 to 155, which were shut down --- and the system ran in this configuration between December 11, 2012 and January 10, 2014 for a total of 120.6 live days, during which $9.6\cdot10^{9}$ T0 and $4.9\cdot10^{8}$ T1 triggers were processed, and $4.1\cdot10^9$ timetraces recorded. The live time of each individual antenna is represented in Fig.~\ref{fig:livetime} for these two periods.

In standard acquisition mode, data quality was monitored through periodic PSD measurements of all detection units present in the DAQ. The antenna trigger rate was also monitored online, and data acquisition was in general halted if the average antenna T0 rate was approaching the maximum 190\,Hz rate (see section~\ref{section:DAQ}), though no strict procedure was defined for data taking interuption. Noisy electromagnetic conditions indeed generate large volumes of data and correspond to reduced DAQ livetime, thus significantly affecting the data quality.

During both East-West and North-South periods, data acquisition was also stopped for various periods of time for maintenance, upgrades or during the one-month annual closure of the site. In 2013, failure of the DAQ room cooling system and a car accident severly damaging the optical fibers network caused several additionnal weeks of interruption of data acquisition. \\

\subsection{Characteristics of the TREND50 electromagnetic environment}
\label{section:envir}
We present in this section the general features of the transient electromagnetic background, as measured by the TREND50 setup. \\

\begin{figure}[t!]
\begin{center}
{\includegraphics[width=6.5cm]{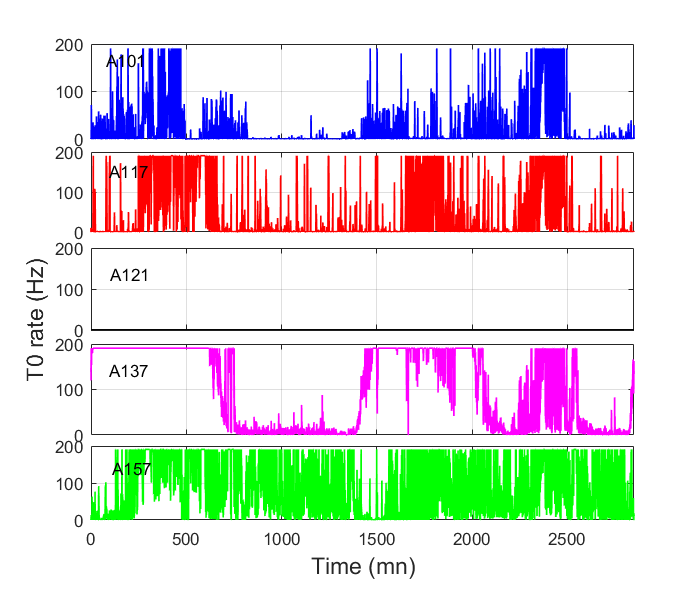}   
\includegraphics[width=6.5cm]{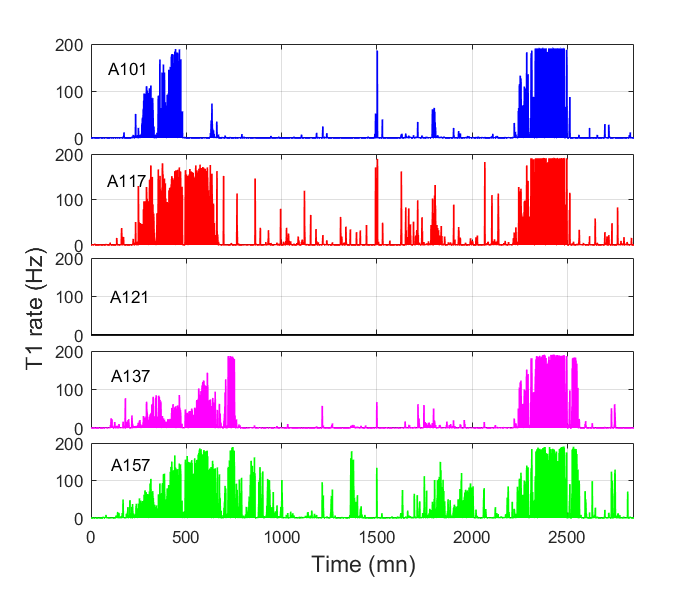}
}
\end{center}
\caption{left: rate of T0 triggers as a function of time for some TREND50 detection units during run R3589 started on March 10, 2012. Right: same for T1 triggers.}
\label{fig:trigrate}
\end{figure}

\begin{figure}[t!]
\begin{center}
{\includegraphics[width=\textwidth]{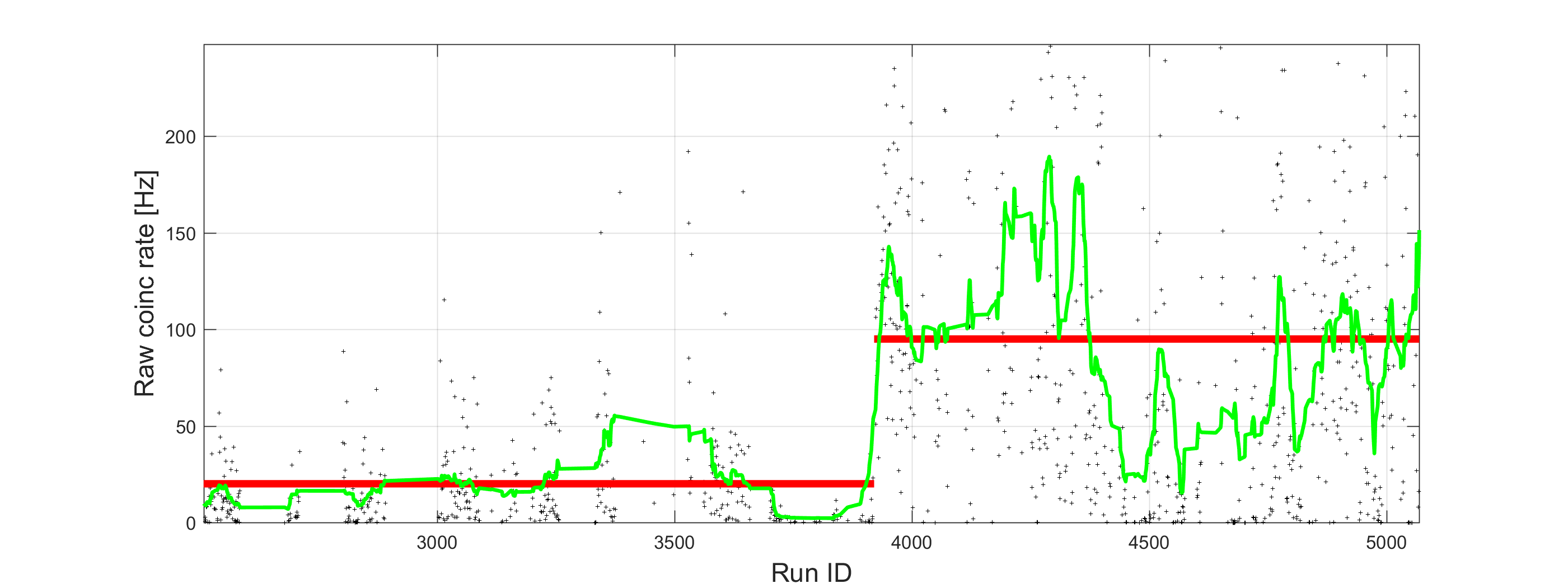} 
}
\end{center}
\caption{run-averaged T1 rate over the whole TREND50 array versus run ID (black dots). In green is displayed the curve smoothed over 30 consecutive runs for the purpose of clarity, and in red the average of the T1 rates over the periods January 13, 2011 - September 16, 2012 and September 17, 2012 - January 10, 2014. Values are 20 and 95\,Hz respectively. }
\label{fig:trigrateevo}
\end{figure}

The first striking feature of the TREND50 data is the very irregular rate of triggers: while T1 rates average 30\,Hz over the three years of data taking, they could shoot up to much larger values for short period of times as illustrated in Fig.~\ref{fig:trigrate} for a typical example. On longer timescales, the data acquisition conditions can be divided in two distinct periods: until september 2012, the total TREND50 T1 rate for a given run was 20\,Hz on average, before rising up to 95\,Hz after that date (see Fig.~\ref{fig:trigrateevo}). This evolution was particularly significant for antennas A101-A120 located in the Eastern part of the array. \\

\begin{figure}[t!]
\begin{center}
{\includegraphics[width=\textwidth]{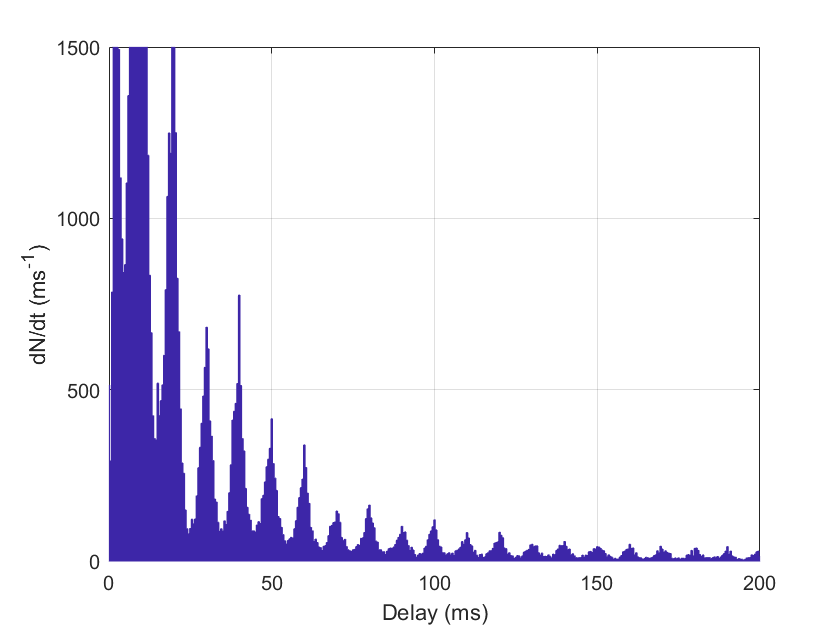}}  
\end{center}
\caption{distribution of the time delays between two consecutive events for data recorded during run R3589. A periodic structure with T=10\,ms clearly appears, signing the fact that many events in these data were generated by 50\,Hz power lines. }
\label{fig:consdelay}
\end{figure}

Another interesting feature is the distribution of time delays $\Delta t$ between consecutive events: a clear periodic structure can be observed, with  $\Delta t$ values clustering around multiples of 10~ms (see Fig.~\ref{fig:consdelay} for an illustration). This indicates that these events most likely originate from the power grid elements surrounding the TREND50 detector, for which discharges may occur --- on HV transformers or power line insulators --- when the current reaches its extremum value, thus generating 100\,Hz-periodic electromagnetic bursts triggering the TREND50 array. \\

\begin{figure}[t!]
\begin{center}
{
\includegraphics[width=6.5cm]{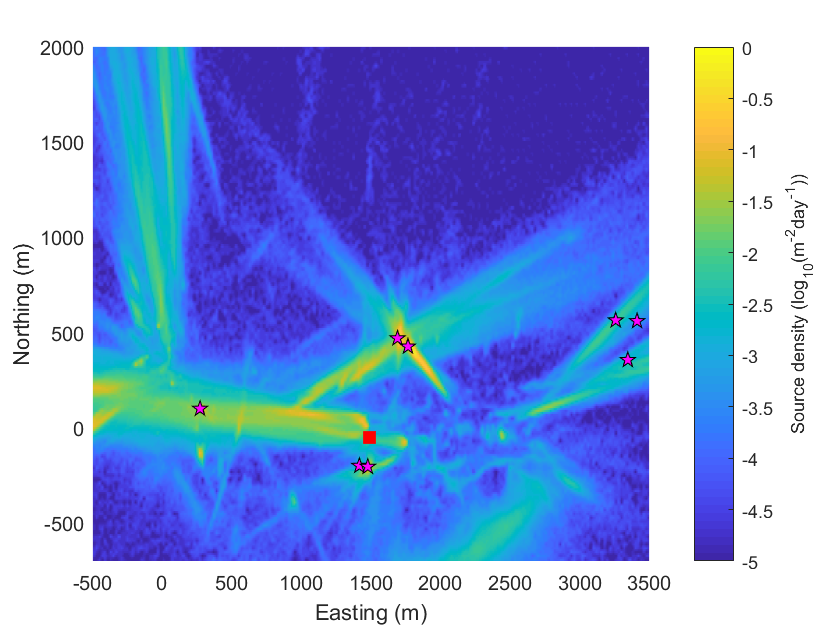}
\includegraphics[width=6.5cm]{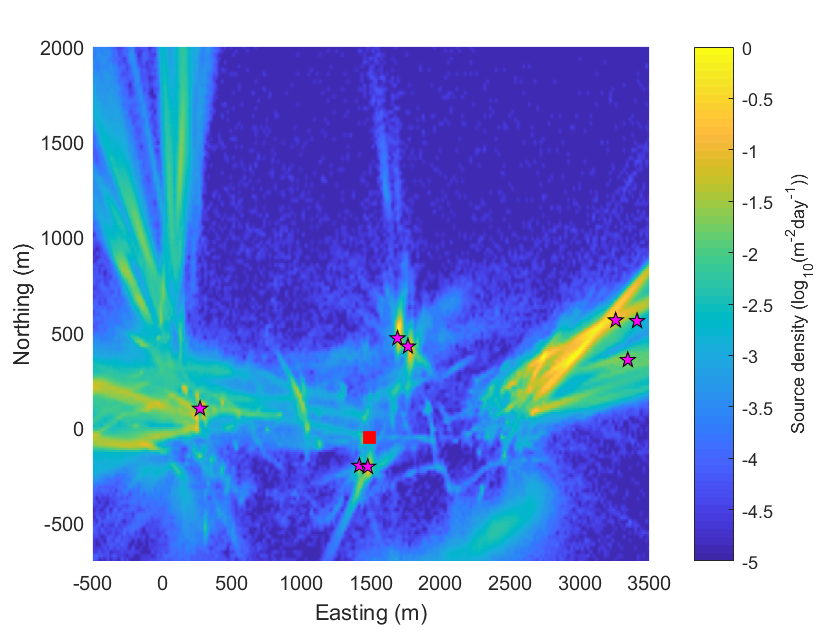}
}
\end{center}
\caption{positions of reconstructed point-sources of TREND50 events for the period until September 2012 (left) and after that date (right). Overled are the locations of the TREND50 DAQ room (red square) and some HV transformers or power line insulators (stars) located in the vincinity of the TREND50 array. Many reconstructed source positions cluster around these positions, confirming that local power system elements constitute sources of background events. The finite extension of the TREND array implies a poor handle on the actual distance to sources, hence generating the long tails observed in the distribution of the reconstructed positions.}
\label{fig:xy}
\end{figure}

Reconstruction of the event sources provides a deeper insight in the TREND50 electromagnetic environment. Because of early selection cuts (see sections~\ref{section:bursts} to~\ref{section:recons} for details), source reconstruction was performed for a small fraction of the total dataset only (see table~\ref{tab:cuteffdata}), following a method presented in section~\ref{section:recons}. Besides, the limited extension of the TREND50 array does not allow for a good handle on the actual distance to sources located outside the array. Yet, it can clearly be observed in Fig.~\ref{fig:xy} that events often cluster around --- or in a direction compatible with --- some specific elements in the TREND50 surroundings. Figure~\ref{fig:xy} also shows that the increased T1 rate observed after September 2012 coincides with a larger number of sources reconstructed in the direction of three electrical transformers located around x=3300\,m.


\begin{figure}[t!]
\begin{center}
{
\includegraphics[width=6.5cm]{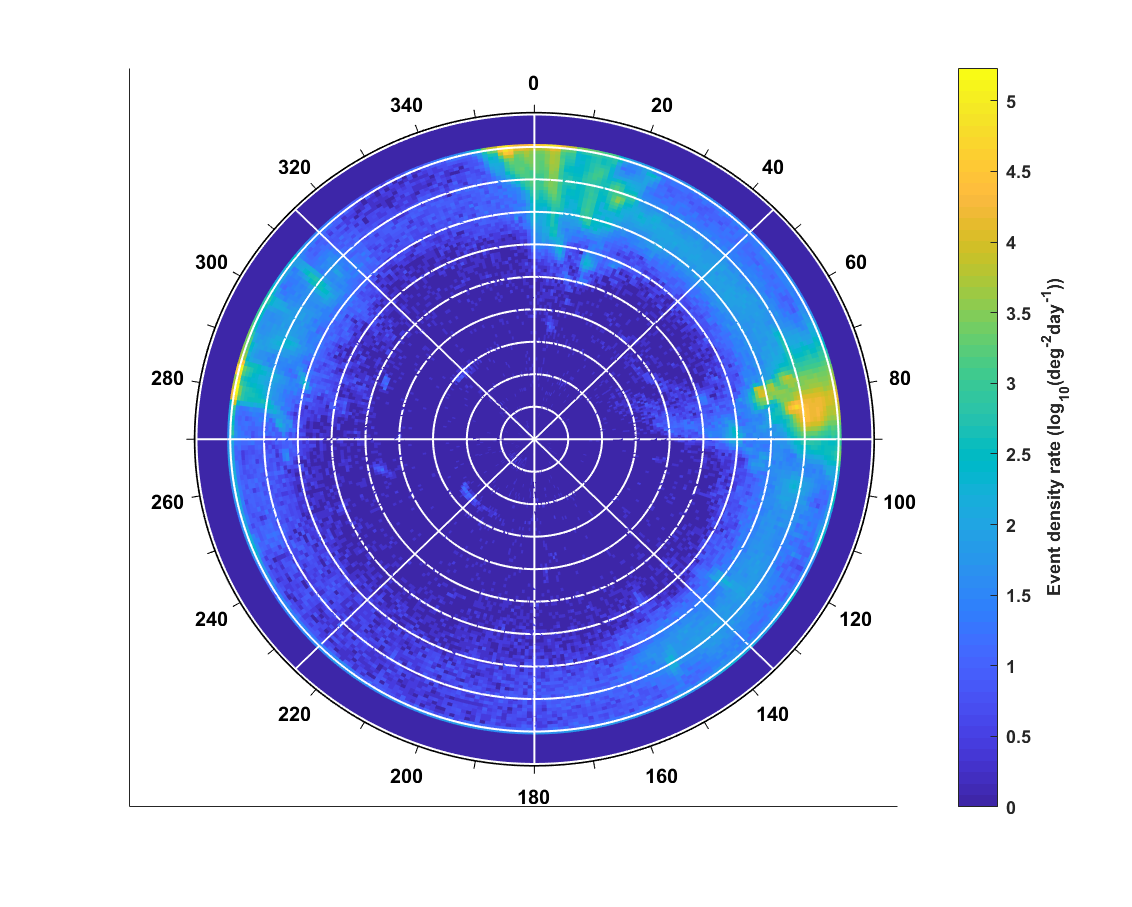}
\includegraphics[width=6.5cm]{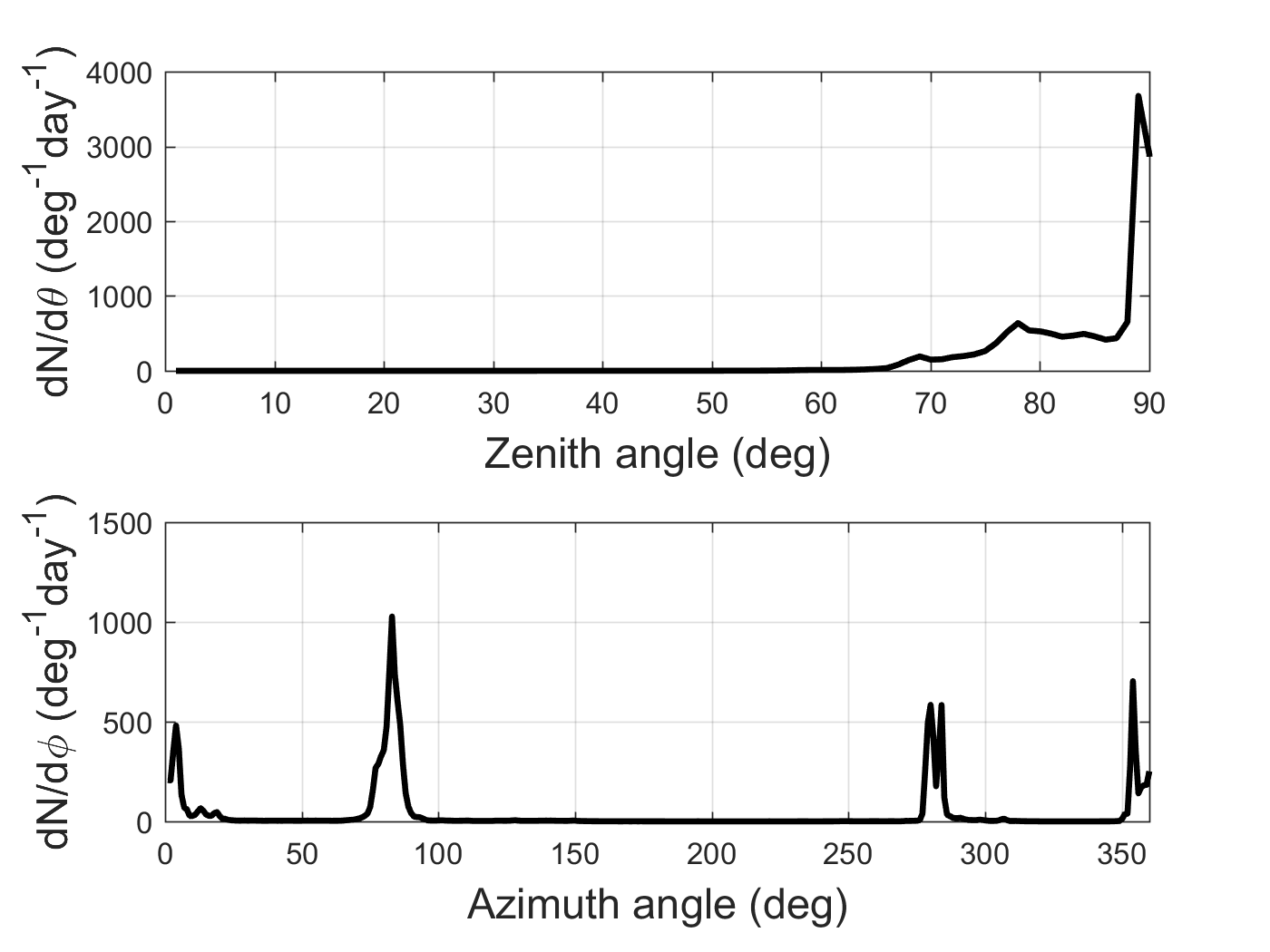}
}
\end{center}
\caption{left: sky distribution of the TREND50 events direction of arrivals. Only events from the East-West period with valid spherical reconstruction and radius of curvature $R \geq $ 3000\,m are shown in this plot. Right: zenith (top) and azimuth (bottom) distribution for the same events.}
\label{fig:distrib}
\end{figure}

For events with sources reconstructed at a distance of $R \geq$ 500\,m from the triggered antennas, a plane wavefront hypothesis is considered as valid, and the direction of origin of the wave can then be discribed by its zenith and azimuth angles ($\theta$,$\phi$). The distribution of these parameters is displayed in Fig.~\ref{fig:distrib} for events with $R\geq$3000\,m. Here again, a large majority of events cluster in specific, fixed directions mostly located along the horizon, eventhough reconstruction errors result in a leakage of this population to smaller zenith angles.  \\

As a conclusion to this section, it clearly appears that, even in a remote location like the TREND site, background sources exist and trigger the antenna array at rates of several tens of Hz, much larger than what is expected for air showers. Yet, a large fraction of these events cluster in time and location, thus giving a good handle for their rejection through a dedicated offline treatment which will be detailed in section~\ref{section:cuts}.

%% file: selection.tex
\section{Air shower identification}
\label{section:selection}

More than 10$^9$ events triggered the TREND50 setup over its 434.9 live days of run, while it can be estimated that only a few thousands cosmic rays with energy higher than 10$^{17}$~eV ---the typical threshold one could expect for an array such as TREND50--- hit the 1.5~km$^2$ array during that period of time. This means that a huge majority of the events detected by TREND50 are generated by background sources, an observation consistent with the results of the previous section. An offline selection procedure was developed in order to discriminate air-shower candidates from the ultra-dominant background events recorded in the experimental data. The different cuts composing this procedure can be classified in two categories: those mainly based on features typical of background signals and aiming at rejecting them ({\it environment cuts}) and those based on the features of the air shower radio signals, aiming at selecting air shower candidates ({\it candidate cuts}).  Before detailing these cuts (section \ref{section:cuts}), we briefly summarize the principles of radio emission by air showers and the associated features of radio signals. The corresponding efficiency on air shower selection is evaluted in section~\ref{section:efficiency} from the simulation dataset presented in section~\ref{section:simu}.

\subsection{Air shower radio emission process}
\label{section:radioemission}
When an ultra-high-energy cosmic ray interacts in Earth's atmosphere, a so-called extended air shower is induced. In 1962, Askaryan already predicted that ambient electrons are dragged along from the medium (mainly due to the Compton scattering process), leading to a net charge excess in the cascade front~\cite{Askaryan:1962}. The time variation of this net excess charge, in combination with Cherenkov effects subsequently leads to coherent radio emission over the typical dimensions of the particle cascade, which is found in the radio frequency range. 

Besides this so-called Askaryan emission, for air showers propagating in Earth's atmosphere a second, leading, emission mechanism is found. Charged leptons inside the cascade are deflected in Earth's magnetic field. This deflection induces a net transverse current in the cascade front. Referred to as the geomagnetic emission, its time variation in combination with Cherenkov effects leads to coherent emission over the typical dimensions of the cascade~\cite{Kahn206,Scholten:2007ky}, as for the net excess charge. The geomagnetic field strength was found to be typically O(10) times larger than the charge excess effect\,\cite{Aab:2014esa}. 

The polarization of the geomagnetic signal is given by the induced transverse current in the shower pointing in the $\vec{v}\times\vec{B}$ direction, where $\vec{v}$ denotes the shower velocity and $\vec{B}$ Earth's magnetic field, while the Askaryan emission shows an observer location dependent radial polarisation pattern, directed towards the shower axis. The combination of both effects leads to a very specific polarization pattern which depends on the positions of the receiving antennas relative to the shower axis. This also applies to the pulses peak amplitudes, which depend on several parameters such as air shower energy, observer position, and for the geomagnetic emission scales with $\sin(\alpha)$, where $\alpha$ denotes the angle of the cascade direction with Earth's magnetic field, as first observed in \cite{CODALEMA:2009}. 
Strong relativistic beaming in the forward direction of the radio emission implies that the footprint of the detectable radiation at ground is limited to an ellipse with minor axis of few hundred meters for non-horizontal trajectories of air-showers. An enhancement of the signal amplitude can be observed on the Cerenkov cone, but it is hardly visible in the TREND frequency range\,\cite{Nelles:2014dja}. Finaly, typical pulse durations are found lasting from several nanoseconds on the Cherenkov cone, up to microsecond timescales for observers located further out.

\subsection{Selection cuts}
\label{section:cuts}

\subsubsection{Noise bursts}
\label{section:bursts}
As illustrated in Fig.~\ref{fig:trigrate}, the antenna trigger rate is very irregular, with large bursts followed by quieter periods. These burst periods represent large volumes of data, but a minor fraction of the acquisition time. Excluding them from the air-shower search sample thus allows removing a large fraction of background events at a moderate cost for air-shower detection efficiency. The following cut was therefore implemented: an event, defined in TREND as a causal coincidence between 4 antennas or more, is rejected if another one occured within a time period of $\pm$200\,ms with at least one detection unit in common.

\subsubsection{Pulse shape}
As mentioned already (see section \ref{section:radioemission}), the transient signal induced at antenna output by an air shower is expected to be brief (pulse width $<$300\,ns at DAQ level) inside the Cherenkov cone. On the other hand, signals from background sources are often significantly longer ($\sim$1\,$\mu$s), and followed or preceded by other transient pulses within a few $\mu$s at most (see Fig. 9 in~\cite{Ardouin:2011}). An offline treatment therefore scans the 5.1\,$\mu$s timetraces and identifies the periods when the signal exceeds the noise level by more than 4$\sigma$. If there is more than one such period in a given timetrace, or if the central period extends beyond 350\,ns, then this timetrace is rejected and the corresponding antenna removed from the event.

\subsubsection{Source reconstruction}
\label{section:recons}
The following stage of the data treatment is the determination of the source of each event. This is performed through a reconstruction procedure detailed and evaluated in~\cite{Ardouin:2011}, and only briefly summarized here. \\

For each event, antenna trigger time values are first adjusted through an offline cross-correlation treatment of the timetraces participating in it. The wave at the origin of the event is then reconstructed through a standard $\chi^2$-minimization procedure, assuming either a plane or a spherical wavefront. In the first case, the adjusted parameter is the vector $\vec{k}$ normal to the wavefront, which then yields zenithal and azimuthal angles $\theta$ and $\phi$ (as defined in section~\ref{section:unit}). In the case of the spherical wavefront, the parameter adjusted during the reconstruction is the position of the source $\vec{X}_0$. It has now become widely accepted that radio shower fronts are hyperbolic\,\cite{Corstanje:2014waa}, even though this is questioned by others\,\cite{CODALEMA:2017}. However, given TREND50 extension and timing resolution, the plane and spherical adjustments implemented here are sufficient, as resuts of simulations show (see Table \ref{tab:cuteff}).

Next, only valid reconstructions are selected: they correspond to a loose cut $\chi^2 /ndf<30$ in both plane and spherical cases ---allowing to keep a large number of events in the following steps of the analysis--- and a linear fit of the relative trigger times vs reconstructed ones yielding a slope in the range [0.9, 1.1] (see Fig. 6 in~\cite{Ardouin:2011}). As direction reconstruction is significantly less precise for 4-fold events, and since some of the other selection cuts have proved to be less efficient with this minimal multiplicity, only events with at least five participating antennas are selected for the subsequent stages of the air-shower search. \\

Two more cuts associated with source reconstruction are applied in the air shower selection procedure:
\begin{itemize}
{\item only events with a radius of curvature $R \geq $3000\,m are selected, where radius of curvature is defined as the distance between $\vec{X}_0$ and the closest antenna participating in the event. The reason for this is that TREND50 is expected to detect air showers with zenith angles larger than 40$^\circ$ in the vast majority of cases (see Fig.~\ref{fig:histos}). For such showers, the $X_{max}$ position ---considered in first approximation as the source of the radio emission--- is indeed further than 3000\,m from the shower core position. On the contrary, a majority of background sources are close enough from the array so that they can be associated with a radius of curvature smaller than this value.}
{\item only events with zenith angle $\theta \leq 80^{\circ}$ are selected because the signal/noise ratio is very unfavorable beyond this value: background events cluster along the horizon (see Fig.~\ref{fig:distrib}), while very few air-showers are expected in this zenith range, in particular because of the reduced sensitivity of the TREND antenna for waves with large zenith values (see Fig.~\ref{fig:leq}). Besides, for a detector deployed over a flat area such as TREND, reconstruction performances significantly degrade for events along the horizon (see Fig. 8 of~\cite{Ardouin:2011}).}
\end{itemize}

\subsubsection{Trigger pattern at ground}
\label{section:bary}
Cosmic ray air showers induce a radio footprint of modest size at ground (see section \ref{section:radioemission}). Besides, the radio emission being at first order linearly polarized, it is expected that, for most showers, all monopolar TREND50 antennas within this footprint receive a comparable signal. On the contrary, in the case of background events generated by distant sources with an elliptical polarization of the wave, some antennas may not trigger if the electromagnetic field happens to be perpendicular to the antenna arm at this specific location. The corresponding events would therefore be characterized by extended illuminated area and/or "holes" in the trigger pattern. Two corresponding cuts are implemented in the TREND offline air shower selection procedure:
\begin{itemize}
{\item it is first requested that the mean distance of the antennas participating in the event to its barycenter position is smaller than 500\,m.}
{\item an event is also rejected if more than one hole in the trigger pattern exists (i.e. at least two non-triggered antennas surrounded by triggered antennas). It must be pointed out here that the polarization and amplitude patterns at ground could certainly be a great tool for background rejection, given the very specific ---and well understood--- features of radio emission by air showers already mentionned in section \ref{section:radioemission}. The peculiar layout of the TREND50 setup (see Fig.~\ref{fig:layout}), combined with the fact that antennas are monopolar, however prevents this selection cut from being optimal for the analysis presented here.}
\end{itemize}

\subsubsection{Neighbouring events}
As detailed already in section~\ref{section:envir}, background events cluster in time (burst periods, see Fig.\ref{fig:trigrate}) and space (fixed background sources location, see Fig.~\ref{fig:distrib}). It is natural to exclude these periods or locations from the air shower search, as they are characterized by disastrous signal/noise ratios. Two corresponding cuts are implemented in the TREND50 air shower search procedure:
\begin{itemize}
{\item an event is rejected if at least another one is found within $\pm$ 30\,s of the candidate's trigger time with a valid reconstruction (as defined in~\ref{section:recons}) and at least 66\% of its antennas in common with the candidate.}
{\item an event is rejected if at least another one is found within $\pm$10\,minutes with a valid reconstruction yielding a radius of curvature $R\geq500$\,m an azimuthal difference $\lvert \Delta \phi \rvert < 10^\circ$, and at least 33\% of its antennas in common with the candidate.}
\end{itemize}

\subsection{TREND air showers simulations}
\label{section:simu}
In order to evaluate the air shower detection efficiency of the whole TREND50 data taking and processing chain, air shower events have been simulated in the most realistic way and processed through the standard TREND50 data analysis chain. The production of these simulated events was as follows:
\begin{enumerate}[(i)]
{\item First, the parameters of the air showers are defined: a fixed value is chosen for the primary particle energy $E_i$ taken in the following set: [5$\cdot 10^{16}$, 7$\cdot 10^{16}$, 8$\cdot 10^{16}$, 1$\cdot 10^{17}$, 2$\cdot 10^{17}$, 3$\cdot 10^{17}$, 5$\cdot 10^{17}$, 7$\cdot 10^{17}$, 1$\cdot 10^{18}$, 2$\cdot 10^{18}$, 3$\cdot 10^{18}$], where energies are given in eV. The direction of origin of the shower ($\theta,\phi$) is drawn randomly from a uniform sky distribution, and the core of the shower $\vec{x}_{core}$ is also a random position in an area $S_{draw}$ surrounding the array, where $S_{draw}$ is defined as a function of $\theta$ and $\phi$, and large enough so that all showers likely to trigger the TREND array fall into $S_{draw}$. The process is repeated $N_{draw}$ times, until $N_{sim}$ trajectories are produced with a minimum of five antennas falling within a distance $d \leq d_{min}$ from the shower axis. Here $d_{min}$ ranges between 500 and 1000\,m depending on $E_i$, and $N_{sim}$ = 10\,000 for energies $E_i\leq1\cdot 10^{17}$~eV and $N_{sim}$ = 3\,000 for $E_i > 1\cdot 10^{17}$~eV. } 

{\item Then the ZHAireS code~\cite{ZHAireS} is used to simulate each shower ($E_i,\theta,\phi,\vec{x}_{core}$) and to compute the E-field transient signals induced by the shower at the TREND50 antenna locations. Simulations are carried out assuming a proton or an iron primary with equal probabilities, and are based on Earth's magnetic field value at the TREND site and the standard US atmosphere.}

{\item The next step consists in simulating the antenna response to these electromagnetic waves using the NEC2 numeric code. At the end of this stage, each air-shower from the simulation set is associated with simulated voltage waveforms at the antenna output and form a so-called {\it simulated event}.}

{\item Then a random time $t^*$ is drawn for each of these simulated events in a time window restricted to TREND50 acquisition periods where calibration data is also available. We then determine from the DAQ monitoring information which detection units participating in the simulated event were active and functionning at this instant, with corresponding processing units also ready to acquire data. Antennas failing this test are discarded from the event. For each remaining antenna, a 50-100\,MHz filtering of the simulated signal is performed and a 200\,MSamples/s digitization applied. The experimental gain value determined from the closest PSD measurement for this specific antenna is then applied to the simulated waveform.
The pretrigger signal of the experimental timetrace recorded at the instant closest in time to $t^*$ is considered as a fair estimate of the noise conditions experienced on the antenna at this instant. It is duplicated to the timetrace length and added to the calibrated waveform, thus forming a simulated timetrace in the standard TREND50 format.
If the maximum amplitude of at least five timetraces from the event exceed the experimental threshold values used in this specific run, then the event is finally inserted in the experimental data set and blinded for the following steps of the analysis. This whole process guarantees that this simulated event corresponds at best to the set of timetraces that would have been recorded on disk in reality, should a shower of energy $E_i$ and direction ($\theta,\phi$) have struck the TREND50 array at location $\vec{x}_{core}$ and instant $t^*$. The process is iterated for each simulated event.}

{\item Finally the TREND50 offline data processing is applied to the data set where the simulated events have been inserted. The processing consists in an implementation of the selection cuts defined in the previous section. At the very end of the chain remains a list of surviving simulated events $N_{sel}$. The results of the offline data processing on the simulated dataset are presented for $E_i=5\cdot10^{17}$\,eV in Table~\ref{tab:cuteff}.
}

\end{enumerate}

It appears from Table~\ref{tab:cuteff} that the two periods result in very contrasted performances. While the East-West period exhibits reasonnable efficiency ---about 40\% of the initial air shower sample pass all implemented cuts---, efficiency drops below 15\% for the North-South period. {\it Environment cuts} are the most penalizing ones for the East-West periods ({\it bursts} or {\it direction neighbours} cuts in particular). Degradation of the electromagnetic environment during the North-South period (see Fig.~\ref{fig:trigrateevo}) translates into much worse performance for the {\it bursts} cut, while this data set efficiency is also affected by {\it signal cuts}, like the {\it pattern} cut, because of a significant degradation of the detector array status over time, with many more noisy or malfunctionning antennas compared to the East-West period.

\ \\
\begin{center}
\begin{tabular}{p{2.4cm}|c|c|c|c}
    \hline
       & \multicolumn{2}{c|}{EW}  & \multicolumn{2}{c}{NS}    \\
    \hline
    Cut    &  Nb  & Survival \%  &   Nb & Survival \%  \\
    \hline
    \hline
    Raw    &   406    &   -        &     209       &    -         \\
    \hline
    Bursts &   309    &   76 $\pm$ 4  &     82       &    39  $\pm$  6     \\
    \hline
    Bad pulses &   279    &   90  $\pm$ 4   &     77       &    94  $\pm$ 6  \\
    \hline
    Valid recons &   272    &   97  $\pm$  2   &    73      &    95 $\pm$  7 \\
    \hline
    Mult$\geq$5 &   272    &   100   $\pm$ 0 &     73       &   100  $\pm$   0  \\
    \hline
    Radius$\leq$ 3000\,m &   263  & 97 $\pm$   2     &     68       &    93 $\pm$  7      \\
    \hline
    $\theta\leq80^\circ$&   246  & 94  $\pm$   3    &     64       &   94 $\pm$   7     \\
    \hline
    Barycenter&   241    &   98  $\pm$ 2  &     60       &    94 $\pm$    6    \\
    \hline
    Pattern&   218    &  90   $\pm$ 4 &     38       &      63$\pm$   12    \\
    \hline
    Direction neighbourgs&   180    &   82   $\pm$ 5 &      31      &    81  $\pm$ 13      \\
    \hline
    Time neighbourgs&   169    &   94 $\pm$ 4  &     30       &    95   $\pm$   8     \\
    \hline
    \hline
    Total           &   169    &   41  $\pm$ 4 &     30       &    14   $\pm$  4     \\

\end{tabular}
\captionof{table}{results of the air-shower selection process for simulated data. See section~\ref{section:cuts} for cut definitions. Figures are computed for events with multiplicity 5 or more and initiated by a  proton and iron primaries in equal proportions with $E_i=5\cdot10^{17}$\,eV. Similar (but often slightly degraded) results are obtained at other energies. Error bars are estimated following a binomial distribution: $\delta f = 2\sqrt{\frac{f(1-f)}{n}}$, with $f$ the event survival fraction and $n$ the number of events before the cut is applied.}
\label{tab:cuteff}
\end{center}

\subsection{Air-shower detection efficiency}
\label{section:efficiency}

\begin{figure}[t!]
\begin{center}
{\includegraphics[width=6.5cm]{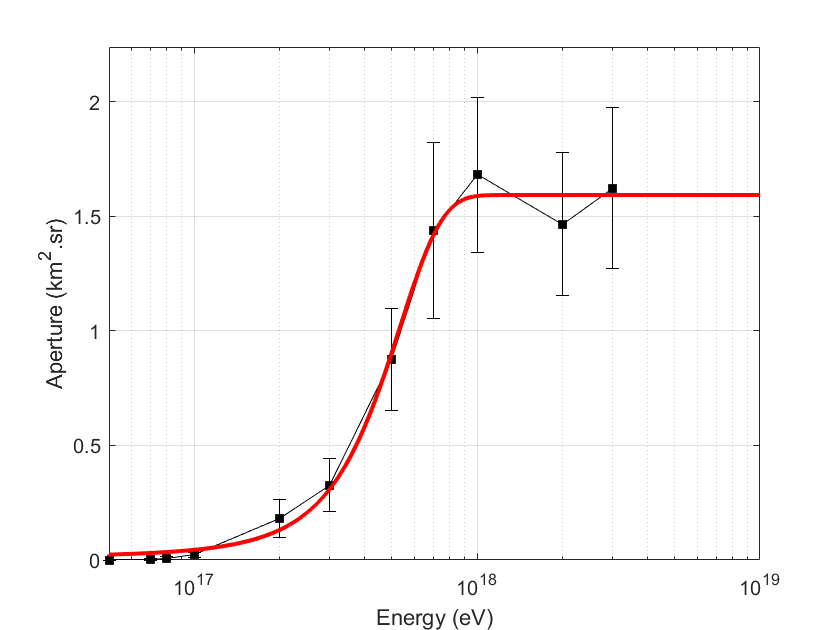}
\includegraphics[width=6.5cm]{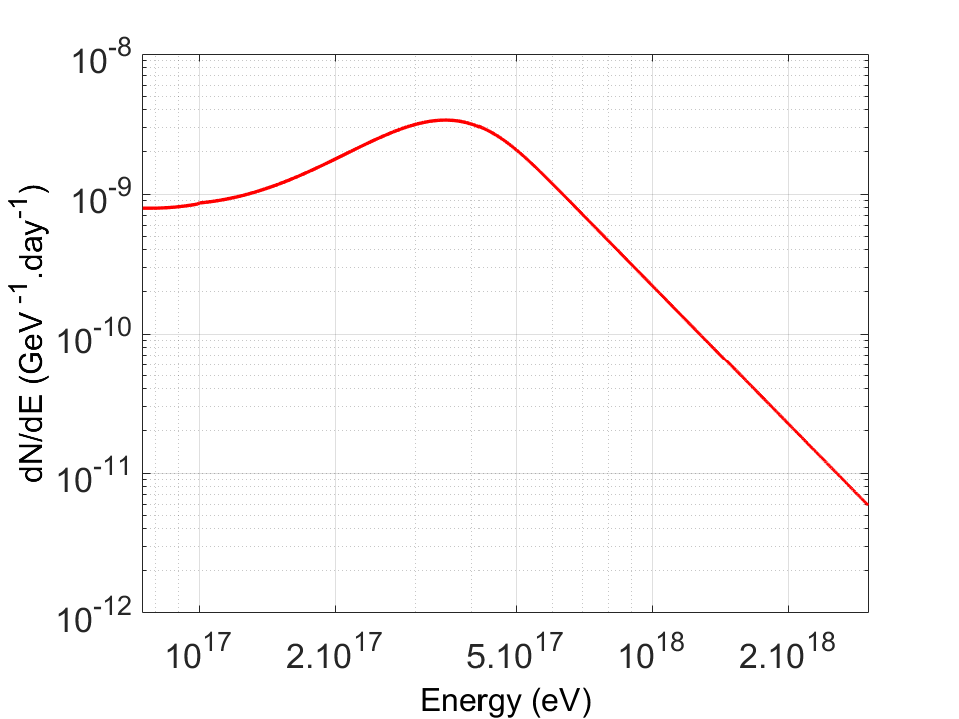}
}
\end{center}
\caption{aperture of the TREND50 array as a function of energy for a fraction of the East-West period, computed from simulations assuming a shared p-Fe composition for the primaries. Right: simulated differential spectrum of cosmic ray detected by TREND, derived from the aperture given in the left plot.}
\label{fig:app}
\end{figure}

The simulation of the TREND50 response to air showers presented in the previous section allows for a straightforward computation of the detector aperture for each energy value $E_i$:

\begin{equation}
\mathcal{A}_{i}=\int_{2\pi}S_{eff}(E_i,\theta,\phi)\cos\theta\sin\theta d\theta d\phi,
\label{eq:ap}
\end{equation}
where the effective area $S_{eff}(E_i,\theta,\phi)$ is given by:
\begin{equation}
S_{eff}(E_i,\theta,\phi)=S_{draw}(E_i,\theta,\phi)\frac{N_{sel}(E_i,\theta,\phi)}{N_{draw}(E_i,\theta,\phi)}.
\end{equation}
The set of $\mathcal{A}_i$ values is then fitted by the analytical function $f(E)$:
\begin{equation}
\label{eq:fitap}
f(E)=a \left( 1+\erf \left( \frac{E-b}{c} \right) \right),
\end{equation}
where $a$, $b$ and $c$ are adjustable parameters. The result is shown in figure~\ref{fig:app}, as well as the simulated differential spectrum of detected events, resulting from the product of $f(E)$ with the differential flux of cosmic-rays $\frac{d^4N}{dE dS d\Omega dt}$ taken from~\cite{augerSpectrum}. 

The number of air shower events expected in duration $\Delta t$ is finally given by:
\begin{equation}
N= \Delta t \int f(E) \frac{d^4N}{dE dS d\Omega dt} dE.
\end{equation}

For the 251 live days of East-West data with valid calibration, we find an expected number of $340\pm60$ air showers detected, and $27\pm8$ air shower events for the 120.6 live days of the North-South period. The statistical errors are estimated by computing the number of events corresponding to a set of apertures $\mathcal{A}_{i}$ shifted by $\pm 1\sigma$ from the values obtained with Eq. \ref{eq:ap}.

\subsection{Systematic effects}
\label{section:errors}
Several biases could affect the computation of the expected number of air showers above described. Below we list the major ones and evaluate their effect for the East-West data set:

\begin{itemize}
\item {\bf Gain:} the gain of the DAQ chain being computed as the ratio of the stationary noise rms at DAQ output to its value at antenna output (see section \ref{gaincomp}), underestimating the noise at antenna level would directly result in a systematical overestimate of the DAQ chain gain. However measurements of the electromagnetic background at the TREND site have shown that the sky noise level is extremely close to what is expected from the Galactic emission \cite{Ardouin:2011}. Besides, the independent calibration cross-check performed on a limited data set results in a   relative difference as low as 3.6\% in average (see Fig. \ref{fig:load}). The systematic overestimate of the DAQ chain gain is therefore of modest or null amplitude. For the purpose of completness, we however calculated that a reduction of the gain by 5\% --- a conservative value--- would lead to 304 air showers detected by the TREND array, instead of 340 determined for the standard DAQ gain values. We associate this value to the systematic uncertainty induced on the number of detected events by the gain computation.

\item{\bf Cosmic rays chemical composition:} experimental results converge towards a variation of the chemical composition of cosmic rays over the TREND energy range, from a  heavy composition around 10$^{17}$\,eV to a lighter one at higher energies \cite{Pierog:2017awp}. Yet, for the purpose of simplicity, a constant composition (50\% proton - 50\% iron) was assumed in our simulations. The impact of this hypothesis was evaluted by computing the response of the TREND array to pure proton and pure iron fluxes. The number of air showers detected by the TREND array is 355 and 325 respectively. 

\item{\bf Air shower radio-emission simulation code:} despite the fact that different simulation codes now converge to very similar results (see \cite{Schroder:2016hrv} for details), we evaluated the response of the TREND array to air showers simulated with CORSIKA+EVA\,\cite{EVA:2012} on a subset of 80 live days of data. The result was found to be statistically compatible with that of ZHaireS. This systematic effect can therefore be neglected.

\end{itemize}

We estimate the systematic error on the number of expected air showers detected by TREND to be the quadratic combination of these different components, thus resulting in a systematic error equal to $340^{+15}_{-39}$ on the number of air showers expected in the 251 live days of East-West period with valid calibration.

%% file: results.tex
\section{Results}
\label{section:results}

\subsection{TREND50 air shower candidates}
\label{section:candidates}
\subsubsection{East-West period}
\label{section:candidatesew}

\begin{figure}[t!]
\begin{center}
{\includegraphics[width=\textwidth]{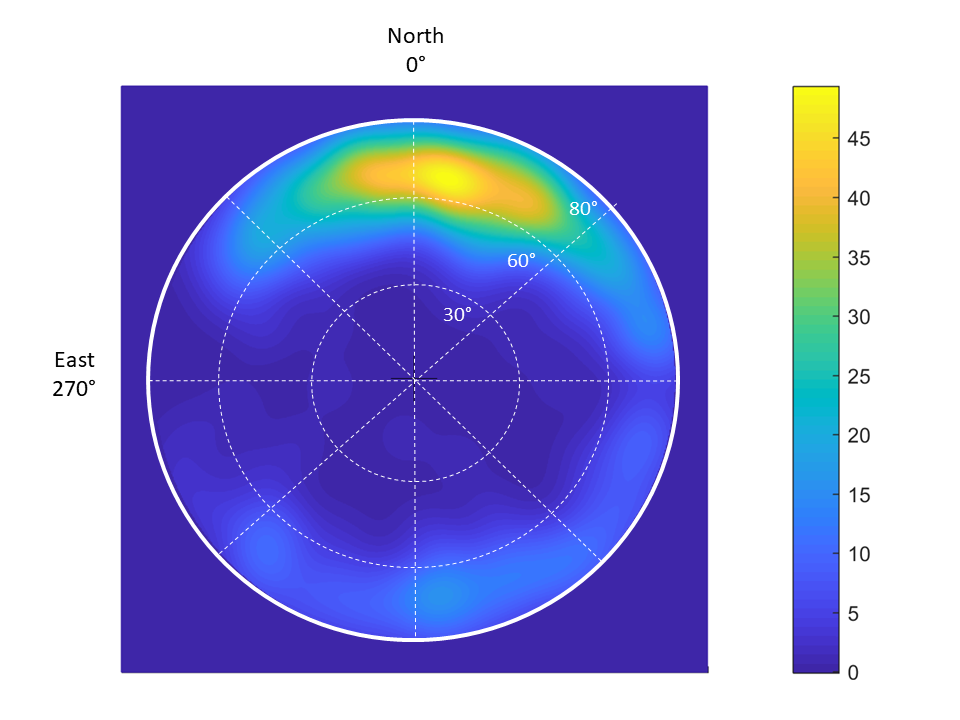}
}
\end{center}
\caption{distribution of the arrival direction for the 564 air shower candidates selected in the East-West experimental data. The distribution is smoothed by the experimental resolution of the direction reconstruction. }
\label{fig:skyplot}
\end{figure}

Out of the $7.3\cdot10^8$ events recorded during the 314.3 live days of the East-West period, a total of 564 air shower candidates survive the selection procedure described in section~\ref{section:cuts} (see Table~\ref{tab:cuteffdata} for details).  The distribution of their reconstructed directions of origin is displayed in Fig.~\ref{fig:skyplot}. It shows a clear excess towards North, as expected for a dominant geomagnetic origin of the radio emission by air showers. One of the candidate is displayed in \ref{candidate} for the purpose of illustration.

\begin{center}
\begin{tabular}{p{2.4cm}|c|c|c|c}
    \hline
       &  \multicolumn{2}{c|}{EW}  & \multicolumn{2}{c}{NS}  \\
    \hline
    Cut    &    Stat        &Surv. \% &    Stat      &Surv. \%\\
    \hline
    \hline
    Raw    &$7.3\cdot10^8$ &      -     &$4.9\cdot10^8$&    -      \\
    \hline
    Bursts   &$6.2\cdot10^7$ &    8.5     &$1.5\cdot10^7$&    3.1      \\
    \hline
    Bad pulses     &$2.9\cdot10^7$ &    46.3     &$7.5\cdot10^6$&    48.8      \\
    \hline
    Valid recons    &$1.9\cdot10^7$ &    67.0     &$3.9\cdot10^6$&    53.0      \\
    \hline
    Mult$\geq$5  & $9.8\cdot10^6$ &    51.2     &$2.8\cdot10^6$&    70.9      \\
    \hline
    Radius$\leq$ 3000\,m   & $3.4\cdot10^6$ &    35.2     &$1.1\cdot10^6$&    37.5      \\
    \hline
    $\theta\leq80^\circ$& $1.0\cdot10^6$ &    30.5     &$3.0\cdot10^5$&    27.9      \\
    \hline
    Barycenter & $9.3\cdot10^5$ &    93.4     &$2.9\cdot10^5$&    97.6      \\
    \hline
    Pattern & $3.4\cdot10^5$ &    37.2     &$2.4\cdot10^5$&    84.5      \\
    \hline
    Direction neighbourgs      & 1400 &    0.4     &      557       &    0.2      \\
    \hline
    Time neighbourgs      & 564 &    41.2    &      118       &    21.2      \\
    \hline
    \hline
    Veto      & - &    -    &      25       &    22.1      \\
    \hline
    \hline

    Total   & 564 & $8.2\cdot10^{-5}$ &   25      &   $4.6\cdot10^{-6}$       \\

\end{tabular}
\captionof{table}{results of the air-shower selection process for experimental data. See section~\ref{section:cuts} for cuts definition, and section~\ref{section:candidates} for the "veto" cut. }
\label{tab:cuteffdata}
\end{center}

\begin{figure}[t!]
\begin{center}
{\includegraphics[width=6.5cm]{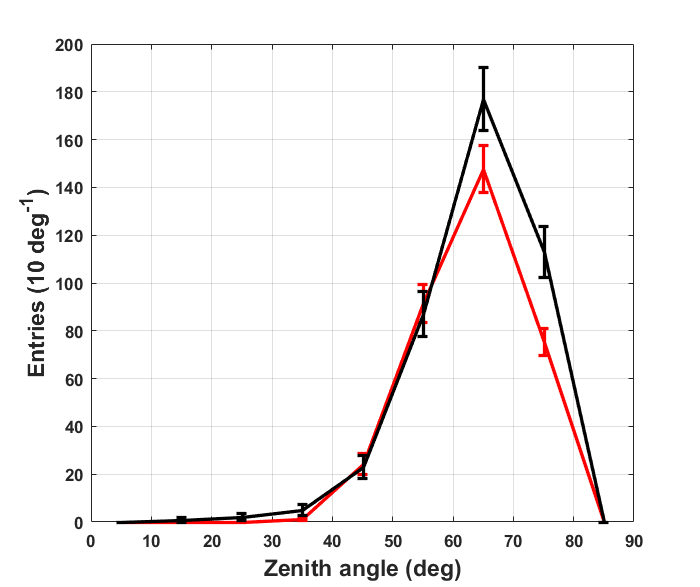}
\includegraphics[width=6.5cm]{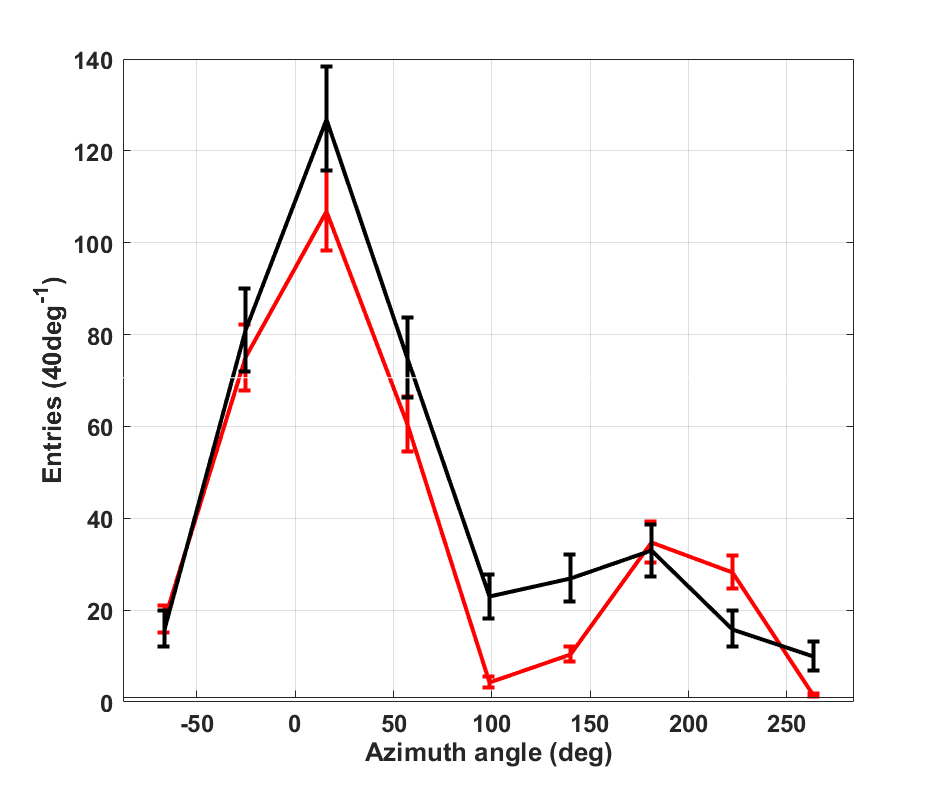}
}
\end{center}
\caption{zenith (left) and azimuth (right) distributions for the 408 air shower candidates selected in the East-West experimental dataset with valid calibration (black) and for the 340 events in the final selection from the simulated data set (red, see section~\ref{section:simu} for details).}
\label{fig:histos}
\end{figure}

Among these 564 candidates, 408 were recorded during the 251 live days with valid calibration, a figure larger than --- but compatible with --- the 340$\pm60^{+15}_{-39}$ air shower events expected during the same period of time according to simulations (see section~\ref{section:efficiency}). The experimental distribution of reconstructed directions, shown in Fig.~\ref{fig:histos}, is also statistically compatible with the simulated set for most directions. An excess of events in the experimental set is however observed for the events closest to horizon and azimuth in the range $-20^\circ \leq  \phi \leq 150^\circ$, a direction where background events are numerous for the East-West period (see Fig.~\ref{fig:distrib} left). This excess is thus understood as a contamination of the final sample of air showers candidates by background events at the $\sim$20\% level. 

This demonstrates that the selection procedure defined for the TREND50 data and detailed in section~\ref{section:cuts} was overall succesfull in discriminating the air shower events from the ultra-dominant background.

\subsubsection{North-South period}
The results are not as good for the North-South dataset, with 118 air shower candidates selected, while 27 were expected according to simulations. The most likely cause for this poor performance is the bad condition of the TREND50 array during most of the North-South period, with many antennas out of order during a large fraction of the data taking time (see Fig.~\ref{fig:livetime}), implying in particular a significant efficiency loss for the selection cut~\ref{section:bary} based on the trigger pattern. This, combined to an increased number of background events due to the degraded electromagnetic environment (see Fig.\,\ref{fig:trigrateevo}), probably resulted in a large contamination of the final sample by background events.  

\begin{figure}[t!]
\begin{center}
{
\includegraphics[width=7.5cm]{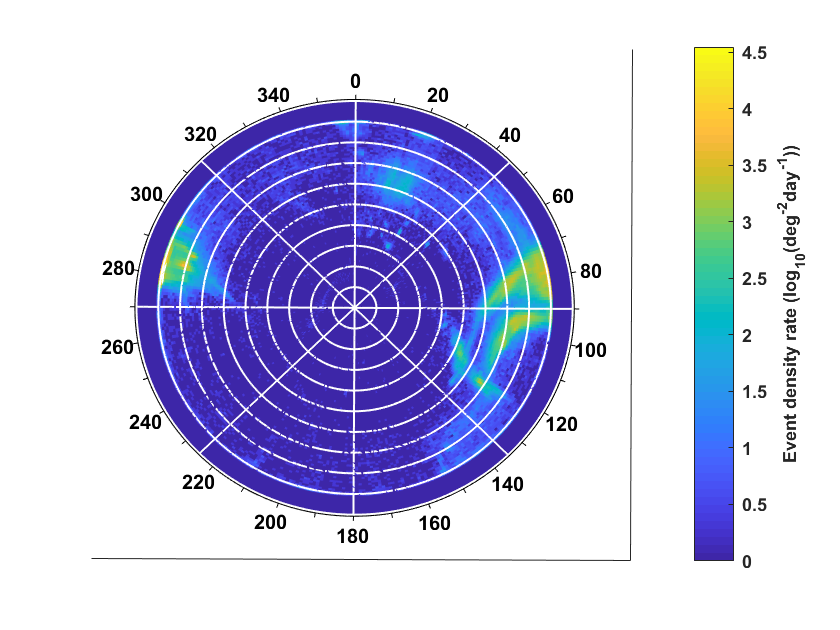}
\includegraphics[width=6.cm]{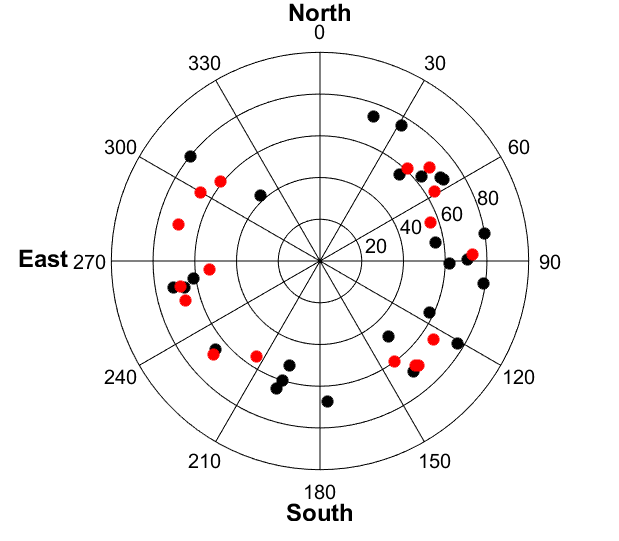}  
}
\end{center}
\caption{left: sky distribution of all events from the North-South period with succesfull point-source reconstruction and radius of curvature $R\geq$3000\,m. Right:  sky distribution of the 25 experimental (black) and 17 simulated (red) air shower candidates from the North-South dataset. }
\label{fig:skyplotNS}
\end{figure}

It however appears that detection units A135 to A138 participate in 93 of these 118 selected candidates, a significant excess over simulations where these detection units are involved in only 10 of the 27 simulated events. Using these peripherical antennas as veto detectors, we reject all events where any of detection units A135 to A138 is involved. The angular distribution of the 25 air shower candidates surviving this {\it veto} cut significantly differs from the background one, and at the same time ressembles the one obtained for the 17 events of the final selection from simulations (see Fig.~\ref{fig:skyplotNS}), even though limited statistics do not allow for firmer statement. 

We can conclude from this paragraph that TREND50 was again able to discriminate air showers from background in this configuration where antennas are oriented along the North-South axis, but with a much reduced efficiency. This also points to the fact that the status and electromagnetic background conditions of the TREND50 array significantly affect the air shower selection procedure.

\subsection{TREND detection efficiency}
Besides computing the efficiency of the offline cuts detailed in Table~\ref{tab:cuteff}, simulations were also used to evaluate TREND's air shower detection efficiency. To do that, we first evaluated the response of an {\it ideal detector}, where all detection units of the TREND50 array are constantly in a running state, and the DAQ system achieves 100\% livetime. This was carried out with a pure proton primary sample, and the corresponding simulated datasets were produced using the same mechanism as detailed in section~\ref{section:simu}, except that instant $t^*$ in step (iv) is replaced by a set of instants $\left\{t_i\right\}$, where each $t_i$ corresponds to optimal conditions for detection unit $i$ (i.e. valid gain, DAQ system up and ready, etc). 

This simulation results in a total of 10768 air showers detected by this {\it ideal detector} during the 251 live days with valid calibration of the East-West period. As 355 air shower events are expected to be detected by the setup in real conditions for proton primaries (see section \ref{section:errors}), the overall detection efficiency of the TREND50 detector can be estimated to 355/10768=3.3\% for this period. 

This is obviously a poor overall performance, but simulations allow for a deeper insight in this result. In particular, when performing the simulation following the standard procedure described in section \ref{section:simu} ---i.e. taking into account the real, instantaneous status of the TREND detector--- it is found that 1110 simulated events would have actually triggered data acquisition and been recorded to disk. Then 355 out of these 1110 events finally pass the offline  air shower selection cuts detailed in section \ref{section:cuts}. The total 3.3\% efficiency may thus be split in two independant parts: a {\it hardware efficiency} of value 1110/10768$\simeq$10\%, and an offline {\it selection efficiency} equal to 355/1110$\simeq$32\%.


The low {\it hardware efficiency} is mostly caused by the moderate reliability of the TREND detector: at any given instant, $\sim$40\% of the detection units could indeed exhibit degraded performances 
or would simply be non-functionnal (e.g. A121 in Fig. \ref{fig:trigrate}, see also Fig. \ref{fig:livetime}). Similarly, the dead time of  processing units was measured to be between 10 and 50\%, depending on noise conditions (see section \ref{section:DAQ}). The {\it selection efficiency}, for its part, is affected by two factors: the peculiar antenna layout, imposed by the necessity to connect to the optical fibers deployed along the 21CMA baselines, does not allow for an efficient mapping of the amplitude pattern at ground, an information that would otherwise be an efficient tool for background rejection. Consequently as much as 35\% of events pass the {\it Barycenter}+{\it Pattern} cuts, while amplitude pattern at ground is potentially a very strong signature for air showers (see section \ref{section:radioemission}). Likewise, measuring only one component of the electric field affects TREND's potential for air shower identification:  
radio emission by air showers indeed exhibit a very peculiar polarization pattern (see section \ref{section:radioemission}), thus providing another strong lever arm for air shower identification which cannot be used in TREND.

Following these observations, the GRANDproto35 experiment\,\cite{GP35:2017,zhang:2015} was proposed. The detection units and front-end electronics are designed to be more robust and stable than TREND. They are associated with a very fast DAQ, allowing for a 100\% livetime for trigger rates up to 10\,kHz according to lab tests\,\cite{GP35:2017}. GRANDProto35 is therefore expected to reach a much better {\it hardware efficiency}. GRANDProto35 detection units will also be equipped with three arms along the East-West, North-South and vertical directions, thus allowing a measurement of the polarization information of the incoming wave. GRANDProto35 {\it selection efficiency} should consequently be improved compared to TREND, though it is not possible at the present stage to give a quantitative estimate for this.

%% file: conc.tex
\section{Conclusion}
We have presented here the results of the TianShan Radio Experiment for Neutrino Detection, a stand-alone, self-triggered radio setup dedicated to the detection of air-showers induced by cosmic rays. 

We have shown that even in the extremely quiet radio environment of the TREND site, our $\sim$1.5\,km$^2$ setup detects background events at a rate of several tens of Hz, orders of magnitude larger than the expected rate of air shower events.

However, the specific  features of both air showers and background events (clustering in time and position in particular) allowed a good rejection of the background events, resulting in a sample of several hundreds of air shower candidates, with an estimated contamination by background around 20\%. 

The TREND air shower detection efficiency is however a few percent only, and dropped to even lower values when the detector status degraded and the environmental noise increased. This modest performance is mostly due to the perfectable status of the TREND50 detector, yielding a hardware trigger efficiency around 10\%, while the cuts implemented for background rejections reached a selection efficiency around 30\% in good electromagnetic conditions. 

The GRANDProto35 project, with an improved DAQ chain compared to TREND and additionnal tools to identify air showers ---measurement of the wave polarization in particular--- aims at achieving a better detection efficiency. It will be our next experimental effort towards the realization of the Giant Radio Array for Neutrino Detection.

%% file: annex.tex
\section{EAS candidate}
\label{candidate}

Below is displayed event 38605 from R3633, recorded by the TREND50 array on April 16, 2012. This is one of the selected EAS candidate.  \\

\begin{figure}[t!]
\begin{center}
{\includegraphics[width=\textwidth]{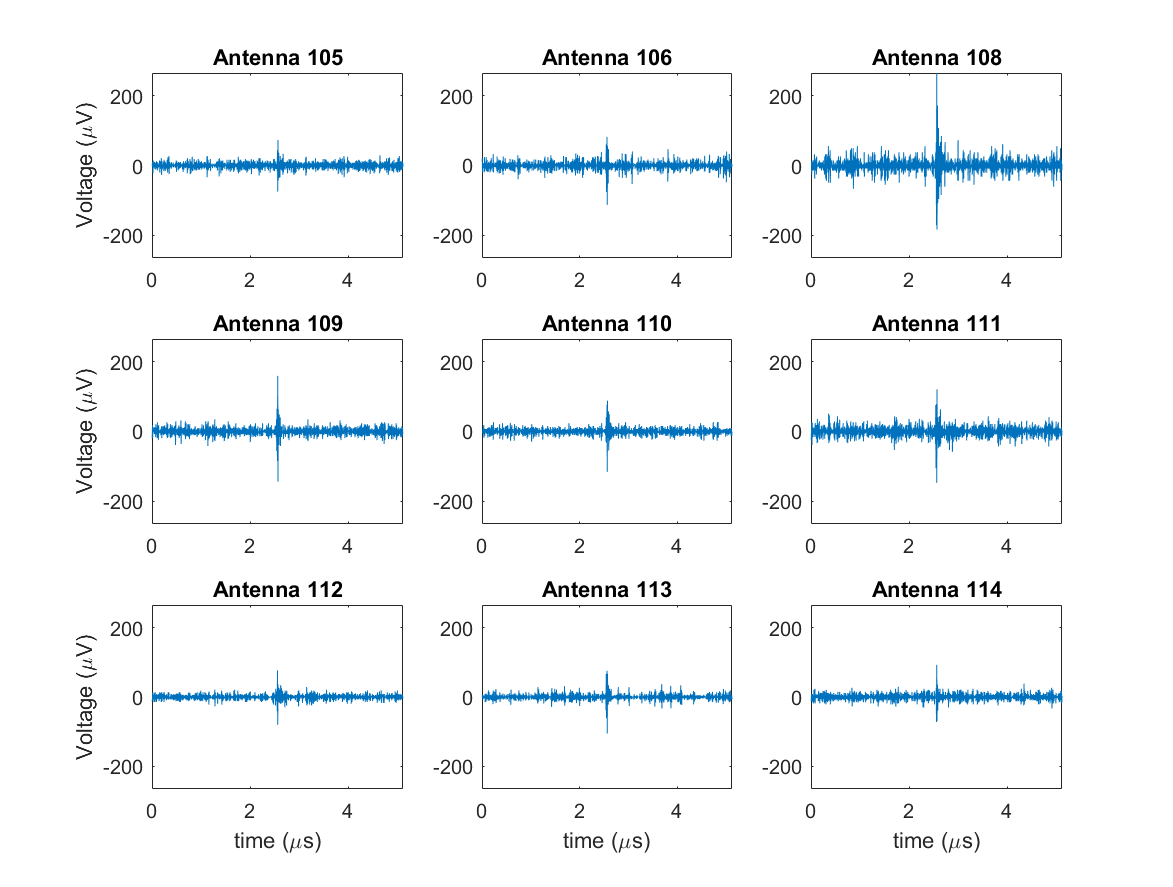}}  
\end{center}
\caption{timetraces of the detection units participating in the event 38605 from R3633.}
\end{figure}

\begin{figure}[t!]
\begin{center}
{\includegraphics[width=0.8\textwidth]{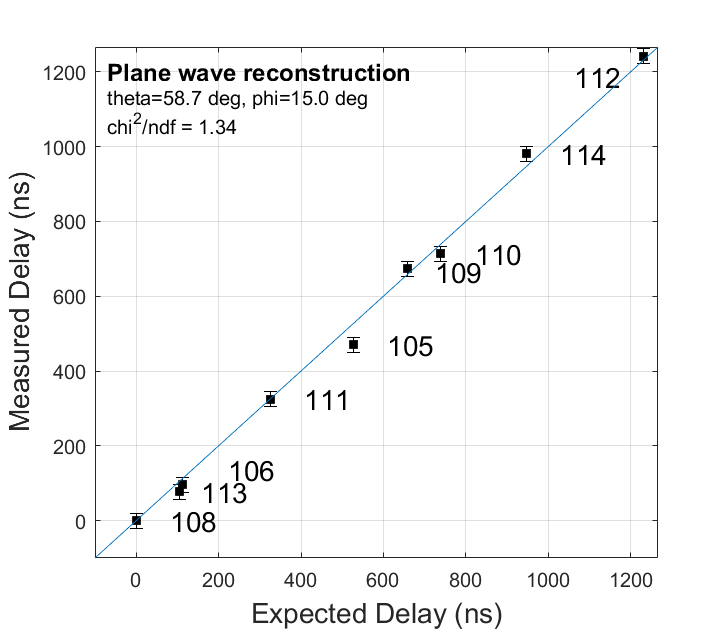}  
\includegraphics[width=\textwidth]{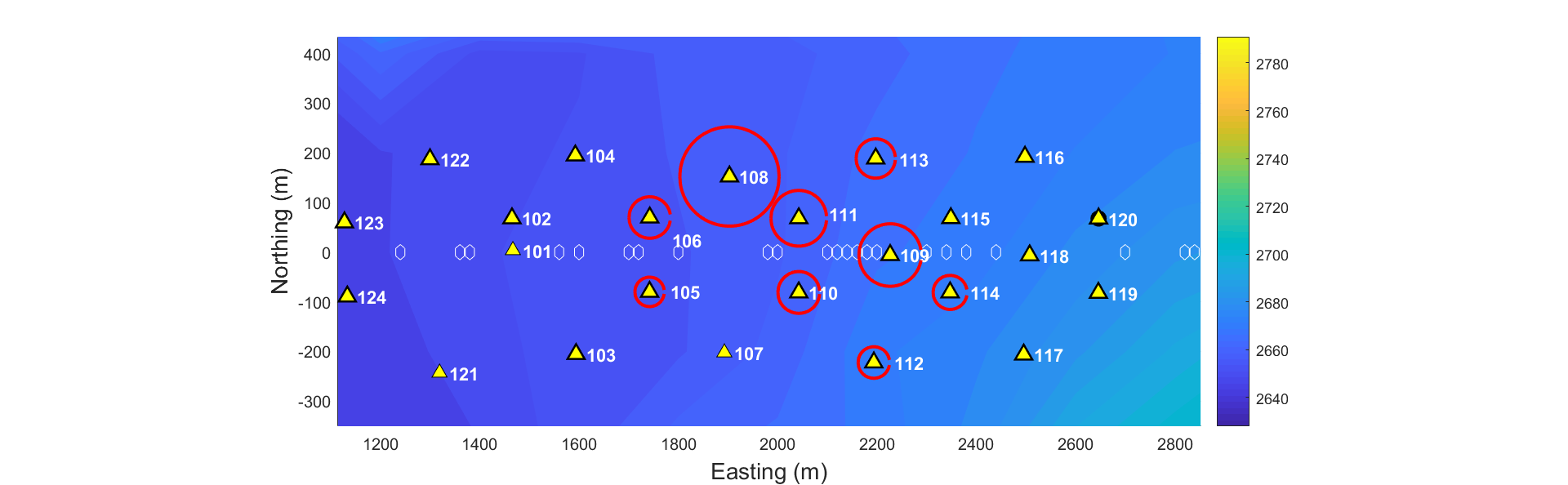}}  
\end{center}
\caption{top: experimental values of the trigger times for the detection units participating in the event 38605 from R3633 as a function of the values reconstructed for a plane wave with zenith and azimuth angles $\theta = 58.7^{\circ}$ and $\phi = 15.0^{\circ}$ respectively. Here the instant of trigger of detection unit 108 is taken as the time reference. Bottom: amplitude pattern of the same event. The radius of the circles are proportionnal to the maximum amplitude of the transient signal on the corresponding detection unit.}
\end{figure}